\documentclass[paper]{JHEP3}
\usepackage{epsfig}
\usepackage{amssymb}
\def\beq{\begin{equation}}
\def\beqn{\begin{eqnarray}}
\def\eeq{\end{equation}}
\def\eeqn{\end{eqnarray}}
\def\abs#1{\left|#1\right|}
\def\HW{{\small HERWIG}}
\def\HWs{{\small HERWIG6}}
\def\HWpp{Herwig{\small++}}
\def\PY{{\small PYTHIA}}

\def\MCatNLO{{\small MC@NLO}}
\def\MCN{{\small MC@NLO}}
\def\IN{{\small IN}}
\def\OUT{{\small OUT}}
\def\bx{\bar x}
\def\dsb{d\bar\sigma}

\def\0t{{\mbox{\bf 0}_\perp}}
\def\tq{\tilde{q}^2}

\def\bk{\bar k}
\def\bp{\bar p}
\def\bs{\bar s}
\def\bt{\bar t}
\def\bu{\bar u}

\newcommand\sss{\scriptscriptstyle}

\newcommand\half{\frac{1}{2}}

\newcommand\tb{\bar{t}}

\newcommand\as{\alpha_{\sss S}}

\newcommand\GenMCatNLO{{\mathcal F}_{\mbox{\tiny MC@NLO}}}
\newcommand\GenMC{{\mathcal F}_{\mbox{\tiny MC}}}
\newcommand\bSigma{\overline{\Sigma}}
\newcommand\mat{{\mathcal M}}

\newcommand\matrmu{{\mathcal M}_\mu^{(3)}}

\newcommand\mato{\overline{\mathcal M}}
\newcommand\matnpo{{\mathcal M}^{(n+1)}}

\newcommand\xMC{|_{\sss {\rm MC}}}
\newcommand\xMCB{\Big|_{\sss {\rm MC}}}
\newcommand\xMCBB{{\Bigg|_{\sss {\rm MC}}}}
\newcommand\clH{{\mathbb H}}
\newcommand\clS{{\mathbb S}}
\newcommand\stepf{\Theta}
\newcommand\Sfun{{\cal S}}
\newcommand\Hone{(H_1)}
\newcommand\Htwo{(H_2)}

\newcommand\fFKS{f_3}
\newcommand\fsi{f_i}
\newcommand\fsa{f_\alpha}

\newcommand\qtt{\tilde{q}^2}
\newcommand\lum{{\cal L}}
\newcommand\mFKS{m}
\newcommand\mFKSi{\mFKS_\pm}
\newcommand\mFKSf{\mFKS_{\fsa}}
\newcommand\yi{y_\pm}
\newcommand\yj{y_{\fsa}}
\newcommand\phii{\varphi_\pm}
\newcommand\phij{\varphi_{\fsa}}
\newcommand\xic{\left(\frac{1}{\xi}\right)_c}
\newcommand\omyid{\left(\frac{1}{1-\yi}\right)_\delta}
\newcommand\opyid{\left(\frac{1}{1+\yi}\right)_\delta}
\newcommand\omyjd{\left(\frac{1}{1-\yj}\right)_\delta}
\newcommand\pt{p_{\sss T}}
\newcommand\kt{k_{\sss T}}

\preprint{
 Cavendish--HEP--10/14\hfill\\
 CERN-TH/2010-212
 }
\title{NLO QCD corrections in Herwig++ with MC@NLO%
\footnote{Work supported in part by the UK Science and Technology 
Facilities Council, and by the Swiss National Science Foundation.}}
\author{Stefano Frixione%
  \thanks{On leave of absence from INFN, Sez. di Genova, Italy}\\
  PH Department, TH Unit, CERN, CH-1211 Geneva 23, Switzerland\\
  ITPP, EPFL, CH-1015 Lausanne, Switzerland\\
  E-mail: \email{Stefano.Frixione@cern.ch}}
\author{Fabian Stoeckli\\
  PH Department, CMG Group, CERN, CH-1211 Geneva 23, Switzerland\\
  E-mail: \email{Fabian.Stoeckli@cern.ch}}
\author{Paolo Torrielli\\
  ITPP, EPFL, CH-1015 Lausanne, Switzerland\\
  E-mail: \email{Paolo.Torrielli@epfl.ch}}
\author{Bryan R.\ Webber\\
  Cavendish Laboratory, %University of Cambridge\\
  J.J. Thomson Avenue, Cambridge CB3 0HE, U.K.\\
  E-mail: \email{webber@hep.phy.cam.ac.uk}}
\abstract{We present the calculations necessary to
  obtain next-to-leading order QCD precision with the \HWpp\ event
  generator using the \MCN\ approach, and implement them for all the
  processes that were previously available from Fortran \HW\ with
  \MCN.  We show a range of results comparing the two implementations.
With these calculations and recent developments in the automatic
generation of NLO matrix elements, it will be possible to obtain NLO
precision with \HWpp\ for a much wider range of processes.
}

\renewcommand\arraystretch{1.1}

\begin{document}

\section{Introduction\label{sec:intro}}
The high-energy collisions taking place at the Tevatron and at
the LHC constitute a severe challenge for any theoretical framework
that aims at predicting them in a quantitative manner.
On the one hand, the typical final-state multiplicity 
can range from a few hundred to a few thousand, with an
average particle energy of the order of a few GeV. At the same 
time, the probability is not negligible to have several (up to 
about ten at the LHC) particles with very large momenta, which
can be used as hard probes for studying the highest-energy 
parton-parton collisions. The task of giving reliable theoretical
descriptions of both these aspects of hadron-hadron collisions
is a very difficult one, and this is why they are usually not
dealt with simultaneously. If one is interested in observables
dominated by the kinematics of multi-particle emissions, 
Parton Shower Monte Carlos (PSMCs) are the method of choice, thanks
to their flexibility and to the fact that they are able to give
a complete (``exclusive'') description of the final states at the
level of measurable hadrons. An alternative is given by the so-called
resummed computations, which organize the perturbative QCD expansion
in terms of the coupling constant $\as$, times the logarithm
(possibly to the second power) of a numerically-large ratio of 
mass scales. The advantage of resummed results over PSMCs is that
the former can (at least in principle) be systematically improved
by including terms less enhanced by logarithms w.r.t. the dominant
ones. However, resummed computations are inclusive, must be performed
observable by observable, are very laborious, and rely on a
fragmentation-type picture to predict hadron-level observables.
As far as the predictions of hard-particle cross sections are concerned,
they are obtained by a perturbative expansion in $\as$ of the relevant 
scattering amplitudes. As in the case of resummed computations, this kind 
of expansion is systematically improvable. However, owing to the complexity
of perturbative calculations in QCD, it is now common to have results
accurate to NLO (next-to-leading order, i.e. one order more than the
one at Born level), while only a handful of results are available to
yet higher orders. It has to be stressed that NLO computations give
sensible results only in the tails of distributions involving large
scales (such as transverse momenta), i.e. for configurations that
have a small probability to happen, or for very inclusive observables
(such as total rates). On the other hand, it is often
the case that interesting physics is characterized by rare events, which 
justify the importance of higher order computations for the Tevatron
and the LHC programmes.

It is clearly desirable to have tools that incorporate the benefits of 
both PSMCs and perturbative computations, without having their drawbacks.
A theoretically-consistent way of matching the two approaches is that
defined by the \MCatNLO\ formalism~\cite{Frixione:2002ik}. \MCatNLO\
requires the modification of the parton-level short-distance cross
sections used in standard NLO computations, achieved through the
insertion of the so-called Monte Carlo (MC) subtraction terms, whose
goal is to remove the double counting that would result by
naively interfacing an NLO result with a PSMC. The MC subtraction
terms can be computed in a process-independent manner, but they are
still dependent on the particular PSMC one adopts for the shower phase.
In other words, each PSMC requires a set of MC subtraction terms.
Although ref.~\cite{Frixione:2002ik} formulated the solution of
NLO-PSMC matching in general terms, practical applications there
and in subsequent papers have been restricted to the choice of
Fortran \HW~\cite{Marchesini:1992ch,Corcella:2001bw,Corcella:2002jc}
as PSMC. Recently, MC subtraction
terms relevant to initial-state emissions have been computed for
\PY\ 6.4~\cite{Torrielli:2010aw}. The aim of this paper is to
present the computations of the MC subtraction terms for the case
of \HWpp~\cite{Bahr:2008pv,Bahr:2008tf}\footnote{Stand-alone versions 
of \MCatNLO\ for \HWpp\ have been implemented for some 
processes~\cite{LatundeDada:2007jg,Papaefstathiou:2009sr,LatundeDada:2009rr}.}.
These terms have then been implemented in the \MCatNLO\ computer 
package. As a consequence, all processes presently implemented in this 
package can be simulated either with Fortran \HW\ or with \HWpp.
We also present here comparisons between the \MCatNLO/\HW\ and
the \MCatNLO/\HWpp~predictions for a few selected observables and
processes in hadronic collisions.

\enlargethispage{10pt}
This paper is organized as follows. In sect.~\ref{sec:mcatnlo} we 
summarize the basic features of the \MCatNLO\ formalism, and in
sect.~\ref{sec:lowmult} we describe their application to the case
of low-multiplicity processes. In sect.~\ref{sec:MCsubt} we give
the general forms of the MC subtraction terms relevant to \HWpp.
Section~\ref{sec:results} presents the comparisons between results
obtained with \HWpp\ and Fortran \HW\ in the context of the \MCatNLO\
approach. Our conclusions are reported in sect.~\ref{sec:concl}.
Technical details specific to \HWpp\ are given in appendices~\ref{sec:HWppzka}
and~\ref{sec:MCsubt2}.

\section{MC@NLO}
\subsection{Generalities\label{sec:mcatnlo}}
The definition of the \MCatNLO\ formalism was given in 
ref.~\cite{Frixione:2002ik}. In a completely general case, one
can write its generating functional as follows:
\beqn
\GenMCatNLO&=&
\sum_{\mu,\mFKS}\int dx_1 dx_2 d\phi_{n+1}\Bigg\{
\GenMC^{(n+1)}\Bigg(\frac{d\bSigma_{\mu|\mFKS}^{(n+1)}}{d\phi_{n+1}}
-\frac{d\bSigma_{\mu|\mFKS}^{\rm\sss(MC)}}{d\phi_{n+1}}\Bigg)
\nonumber\\*&&\phantom{\sum_{\mu,\mFKS}\int dx_1 dx_2 d\phi_{n+1}\Bigg\{}
+\GenMC^{(n)}\Bigg(\frac{d\bSigma_{\mu|\mFKS}^{\rm\sss(MC)}}{d\phi_{n+1}}
+\frac{d\bSigma_{\mu|\mFKS}^{(n)}}{d\phi_{n+1}}\Bigg)\Bigg\}\,.
\label{genFdef}
\eeqn
Here, we assume that the hard process has $2\to n$ and $2\to n+1$ 
kinematic configurations at the Born and real-emission level respectively;
the index $\mu$ runs over all real-emission processes; the role of
the index $\mFKS$ will be clarified in the following.
We have denoted by $\GenMC^{(k)}$ the generating functional of the
Parton Shower Monte Carlo (PSMC) the NLO computation is matched to,
where the index $k$ indicates that the initial condition for the
shower is given by a $2\to k$ partonic configuration. In the case 
\mbox{$k=n+1$}, this configuration coincides with that associated with
the phase-space point \mbox{$(x_1,x_2,\phi_{n+1})$}, and is called
real-emission or event kinematics. When $k=n$, it is obtained from
the event kinematics by means of a projection, which is dictated by
the structure of the underlying NLO computation, and is called
counterevent kinematics. The \MCatNLO\ formalism requires the
NLO cross section be computed by means of the so-called
FKS subtraction~\cite{Frixione:1995ms}\footnote{Strictly speaking,
\MCatNLO\ may be formulated in terms of any other subtraction
formalism, but in practice this has never been done.}. The basic idea 
of the method is the following: for a given $2\to n+1$ real-emission
process $\mu$, one introduces a set of arbitrary functions
$\Sfun_{\mu|\mFKS}$ (called $\Sfun$ functions) which obey the constraint:
\beq
\sum_{\mFKS} \Sfun_{\mu|\mFKS} = 1\,.
\label{Ssum}
\eeq
Each value of the index $m$ is equivalent to the labels of two 
(strongly-interacting) particles entering the process $\mu$, whose
collinear configurations will cause the real-emission
matrix element to diverge, and therefore the sum in eq.~(\ref{Ssum}) 
can be interpreted as running over all possible singular regions
of process $\mu$. Roughly speaking, the $\Sfun$ functions
are constructed in such a way that, in the singular region associated
with $\mFKS$, all $\Sfun_{\mu|\mFKS^\prime}$ with $\mFKS^\prime\ne\mFKS$ 
are equal to zero, and \mbox{$\Sfun_{\mu|\mFKS}=1$}\footnote{The case
of soft singularities is technically more involved, but does not present
any difficulty of principle. The interested reader is referred to the 
original publication  and to ref.~\cite{Frederix:2009yq} for further 
details, knowledge of which is irrelevant here.}. This leads one to introduce 
the quantities
\beq
\matnpo_{\mu|\mFKS} = \Sfun_{\mu|\mFKS}\matnpo_\mu\,,
\label{realME}
\eeq
where $\matnpo_\mu$ is the real-emission matrix element squared for 
the process $\mu$.
Owing to the properties of the $\Sfun$ functions, each $\matnpo_{\mu|\mFKS}$
has at most one soft and one collinear singularity, which can therefore
be subtracted in an essentially trivial manner. In other words, the
$\Sfun$ functions effectively achieve a partition of the phase-space,
and the index $\mFKS$ labels the members of this partition.
Using eq.~(\ref{realME}), one introduces
\beq
\frac{d\bSigma_{\mu|\mFKS}^{(n+1)}}{d\phi_{n+1}}=
\lum\,\matnpo_{\mu|\mFKS}\,,
\label{bSigma}
\eeq
$\lum$ being the luminosity (i.e. the product of the PDFs),
and writes the differential cross section for an observable $O$
at the NLO as follows:
\beq
\frac{d\sigma}{dO}=\sum_{\mu,\mFKS}\int dx_1 dx_2 d\phi_{n+1}\Bigg\{
\delta(O-O_{n+1})\frac{d\bSigma_{\mu|\mFKS}^{(n+1)}}{d\phi_{n+1}}
+\delta(O-O_{n})\frac{d\bSigma_{\mu|\mFKS}^{(n)}}{d\phi_{n+1}}\Bigg\}\,,
\label{NLOxsec}
\eeq
where $O_k$ is the definition of the observable $O$ in terms of
$2\to k$ kinematic configurations. Eq.~(\ref{NLOxsec}) implicitly
defines the $n$-body contribution to the cross section,
$\bSigma_{\mu|\mFKS}^{(n)}$, which also enters the \MCatNLO\ generating
functional, eq.~(\ref{genFdef}). The notation used here is extremely
compact, and in particular one may wonder why an $n$-body contribution
is written in terms of the $n+1$-body phase space. We point out that
all the relevant computations are given in great details in
refs.~\cite{Frixione:1995ms,Frederix:2009yq}, but that the key point
here is the observation that all ingredients necessary for the 
implementation of the \MCatNLO\ formalism are already present in
the computation of the corresponding NLO cross section, except
for the so-called Monte Carlo subtraction terms 
$\bSigma_{\mu|\mFKS}^{\rm\sss(MC)}$.

As clarified in ref.~\cite{Frixione:2002ik}, the MC subtraction terms
depend on the PSMC one adopts for showering the hard events. They can
be computed by formally expanding the PSMC results to the same order
in $\as$ as the corresponding NLO contribution to the parton-level 
cross section (i.e. that of the real-emission matrix elements).
Furthermore, their structures are such that all non trivial process-specific 
information is contained in the Born matrix elements. These matrix elements 
are multiplied by kernels whose analytic forms depend solely on the 
shower variables used by the PSMC to generate the elementary branchings,
and on the identities of the partons involved in such branchings.
The MC subtraction terms can therefore be computed in a 
process-independent manner, and subsequently used in eq.~(\ref{genFdef})
to simulate a given process whose NLO short-distance cross section
has been computed previously.

By construction, the MC subtraction terms cancel locally the divergences
of the NLO short-distance cross section. This implies that the quantities
that multiply $\GenMC^{(n+1)}$ and $\GenMC^{(n)}$ in eq.~(\ref{genFdef})
are separately finite everywhere in the phase space. This allows
one to unweight them, and in this way to associate a constant weight
with the corresponding kinematic configurations, as is possible with 
LO-based PSMC; these configurations are called $\clH$ and $\clS$ events
for $(n+1)$-body and $n$-body final states respectively. We point out
that a similar unweighting cannot be performed in the context of a 
pure-NLO computation, eq.~(\ref{NLOxsec}), owing to the fact that the
two contributions in the integrand there are associated with different
kinematics except on a zero-measure subset of the phase space.

So far the MC subtraction terms have been computed for Fortran 
\HW~\cite{Frixione:2002ik,Frixione:2003ei,Frixione:2005vw}
and (partly) for \PY~\cite{Torrielli:2010aw}. In the following we discuss 
applications to low-multiplicity processes and computations specific to \HWpp.

\subsection{Applications to low-multiplicity processes\label{sec:lowmult}}
The structure of eqs.~(\ref{genFdef}) and~(\ref{NLOxsec}) is suited
to the description of processes with arbitrarily large $n$.
On the other hand, all practical applications have been restricted 
to cases for which $n=1$ or $n=2$ so far. We stress that this is the
number of strongly-interacting particles at the Born level. Accompanying
particles such as electroweak bosons or leptons, and decay products of
the primary particles, may effectively enlarge $n$, but this does not
affect the core of the NLO computation, which is our concern here.

When $n$ is a small, simplifications are possible (but not mandatory)
in the structure of eqs.~(\ref{genFdef}) and~(\ref{NLOxsec}): for a given
$\mu$, one may group together several $\Sfun_{\mu|\mFKS}$ contributions,
owing to the fact that the corresponding kinematic configurations are
trivially related to each other. In order to be more precise, we shall
give here two specific examples: we consider the two partonic processes
\beqn
g(p_1)+g(p_2)&\longrightarrow& t(k_1)+\tb(k_2)+g(k_3)
\label{ggttg}
\\
u(p_1)+b(p_2)&\longrightarrow& t(k_1)+d(k_2)+g(k_3)
\label{ubtdg}
\eeqn
which are part of the real-emission contributions to top-pair and
single-top ($t$ channel) production. We label the initial-state
partons with momenta $p_1$ and $p_2$ by $+$ and $-$ respectively,
and final-state particles with momenta $k_i$ by $f_i$. For the two
processes in eqs.~(\ref{ggttg}) and~(\ref{ubtdg}) there are four 
independent $\Sfun$ functions; the corresponding four values of
$\mFKS$ are:
\beq
\mFKS=(\fFKS,+)\,,\;\;\;
\mFKS=(\fFKS,-)\,,\;\;\;
\mFKS=(\fFKS,f_1)\,,\;\;\;
\mFKS=(\fFKS,f_2)\,,
\label{sings}
\eeq
which are associated with the configurations in which the final-state
gluon $g(k_3)$ may become collinear to one of the initial-state partons
($\mFKS=(\fFKS,+)$ or $\mFKS=(\fFKS,-)$), or to one of the final-state 
particles
($\mFKS=(\fFKS,f_1)$ or $\mFKS=(\fFKS,f_2)$). At this point, one observes that,
in any of these collinear configurations, $p_1$ is back-to-back with
$p_2$, and $k_1$ is back-to-back with $k_2$. In turn, this implies
that when e.g. $g(k_3)$ is collinear with $g(p_1)$, it is {\em anti}collinear
to $g(p_2)$. In the FKS subtraction formalism, each $\Sfun_{\mu|\mFKS}$ 
dictates the choice of the integration variables directly related to the
subtraction procedure. In particular, one of these variables is the
angle between the two particles identified by the label $\mFKS$.
It is obvious that, if one given angular variable is suited to
performing a subtraction in a collinear region, it will be as well suited
to performing it in an anticollinear region. This implies that a single
angular variable can be used to deal simultaneously with contributions 
$\mFKS=(\fFKS,+)$ and $\mFKS=(\fFKS,-)$, and another one to deal simultaneously
with $\mFKS=(\fFKS,f_1)$ and $\mFKS=(\fFKS,f_2)$.

One further observes that the contributions $\mFKS=(\fFKS,f_1)$ and
$\mFKS=(\fFKS,f_2)$ to eq.~(\ref{ggttg}), and the contribution 
$\mFKS=(\fFKS,f_1)$ 
to eq.~(\ref{ubtdg}), do not correspond to collinear matrix-element
singularities, but only to soft ones. In the FKS procedure,
the choice of the angular variable does not play any role in the
subtraction of soft singularities, since such a subtraction is dealt
with by a variable which is essentially the energy of the parton becoming
soft. This implies that, in the case of eq.~(\ref{ggttg}), the contributions
$\mFKS=(\fFKS,f_1)$ and $\mFKS=(\fFKS,f_2)$ can be treated together
with those $\mFKS=(\fFKS,\pm)$. In the case of eq.~(\ref{ubtdg}), on the 
other hand, the contribution $\mFKS=(\fFKS,f_1)$ can be treated simultaneously 
with either $\mFKS=(\fFKS,\pm)$, or with $\mFKS=(\fFKS,f_2)$. Although either 
choice is possible, in ref.~\cite{Frixione:2005vw} the latter, more logical, 
option was adopted.

The bottom line is that, owing to the simplified kinematics of $2\to 1$
or $2\to 2$ processes, some of the contributions in the sum over $m$ 
in eqs.~(\ref{genFdef}) and~(\ref{NLOxsec}) can be dealt with together.
In such cases, it is still convenient to use the notation introduced 
in eqs.~(\ref{genFdef}) and~(\ref{NLOxsec}), but $\mFKS$ will need
be understood as a set of pairs of particle labels, rather than as
a single pair. For the specific examples considered above, we shall have
\beqn
&&\mFKS=\Big\{(\fFKS,+),(\fFKS,-),(\fFKS,f_1),(\fFKS,f_2)\Big\}\,,
\label{ttblabels}
\\
&&\mFKS=\Big\{(\fFKS,+),(\fFKS,-)\Big\}\,,\;\;\;\;
  \mFKS=\Big\{(\fFKS,f_1),(\fFKS,f_2)\Big\}\,,
\label{stlabels}
\eeqn
for the processes in eqs.~(\ref{ggttg}) and~(\ref{ubtdg}) respectively.
Therefore, in the case of top-pair production, the sums over $\mFKS$ 
in eqs.~(\ref{genFdef}) and~(\ref{NLOxsec}) contain only one term and,
thanks to eq.~(\ref{Ssum}), it is actually not necessary to introduce
the $\Sfun$ function in order to perform the computation. The case of
single-top is only slightly more complicated. The sums over $\mFKS$ 
contain two contributions; in the original publication,
ref.~\cite{Frixione:2005vw}, the two values of $\mFKS$ in 
eq.~(\ref{stlabels}) have been denoted by \IN\ and \OUT\ respectively.

We conclude this section by pointing out that, regardless of the number
of final-state particles $n$, contributions  $\mFKS=(\fsi,+)$ and 
$\mFKS=(\fsi,-)$ for a given particle label $\fsi$ can be always treated 
simultaneously in the context of the FKS subtraction -- see appendix~E of 
ref.~\cite{Frederix:2009yq} for a discussion on this matter.

\subsection{MC subtraction terms for Herwig++\label{sec:MCsubt}}
As discussed previously, the MC subtraction terms 
$\bSigma_{\mu|\mFKS}^{\rm\sss(MC)}$ are determined by formally
expanding the results of the PSMC in $\as$, and by keeping the
contribution that has the same power in $\as$ as the NLO cross section
(which we shall denote by ${\cal O}(\as^{b+1})$).
This contribution can be written in full generality as follows:
\beq
d\sigma\xMCB=\sum_{\mu,m}\sum_{L\in m}\sum_{l}
d\sigma_{\mu}^{(L,l)}\xMCB\,.
\label{MCatas}
\eeq
The sum over $\mu$ has precisely the same meaning (and range) as
that in eqs.~(\ref{genFdef}) and~(\ref{NLOxsec}), since the order
in $\as$ at which we are working is the same as in those equations,
and hence the partonic processes that contribute to the cross sections
are the same. The sums over $m$ and $L$ can be understood as follows.
At ${\cal O}(\as^{b+1})$, the PSMC gets contributions from the
diagrams that can be constructed by attaching to Born-level diagrams
all possible $1\to 2$ (QCD) branchings stemming from external legs.
These diagrams are therefore a subset of those contributing to 
real-emission matrix elements, and in particular are all diagrams
that may give rise to collinear and/or soft singularities. Hence,
they can be identified by means of the index $\mFKS$ introduced
previously. According to the discussion in sect.~\ref{sec:lowmult},
each $\mFKS$ is a set of pairs of particle labels, with one (in the 
straightforward implementation described in sect.~\ref{sec:mcatnlo})
or more (in the case simplifications are possible) elements.
Either way, the index $L$ in eq.~(\ref{MCatas}) runs over the 
elements of a given $\mFKS$. This implies that each element
in the double sum over $\mFKS$ and $L$ in eq.~(\ref{MCatas}) identifies
unambiguously one ``branching'' in a real-emission configuration.
Note, therefore, that at fixed $\mu$ there is a one-to-one correspondence
between $L$ and a particle in the underlying Born-level process;
such a particle may thus be referred to as the branching leg.
Finally, for each branching leg the PSMC may give rise to different
showers, depending on the colour partner of the branching leg. The sum
over the choice of colour partners is that over the index $l$ 
in eq.~(\ref{MCatas}); note that, at fixed $L$, a given particle can
play the role of colour partner more than once, depending on the
colour flows of the underlying Born process.

The similarities between Fortran \HW\ (denoted henceforth by \HWs)
and \HWpp\ are such that the forms of the cross sections 
$d\sigma_{\mu}^{(L,l)}\xMC$ are the same as those given in 
ref.~\cite{Frixione:2005vw}. They read:
\beqn
d\sigma_{\mu}^{(+,l)}\xMCB &=&
\frac{1}{z_+}
f_a^{\Hone}(\bx_{1i}/z_+)f_b^{\Htwo}(\bx_{2i})\,
d\hat{\sigma}_{\mu}^{(+,l)}\xMCB d\bx_{1i}\,d\bx_{2i}\,,
\label{eq:spl}
\\
d\sigma_{\mu}^{(-,l)}\xMCB &=&
\frac{1}{z_-}
f_a^{\Hone}(\bx_{1i})f_b^{\Htwo}(\bx_{2i}/z_-)\,
d\hat{\sigma}_{\mu}^{(-,l)}\xMCB d\bx_{1i}\,d\bx_{2i}\,,
\label{eq:smn}
\\
d\sigma_{\mu}^{(\fsa,l)}\xMCB &=&
f_a^{\Hone}(\bx_{1f})f_b^{\Htwo}(\bx_{2f})\,
d\hat{\sigma}_{\mu}^{(\fsa,l)}\xMCB d\bx_{1f}\,d\bx_{2f}\,,
\label{eq:sfA}
\eeqn
where, consistently with sect.~\ref{sec:lowmult}, we have denoted by
$+$ and $-$ the two initial-state branching legs, and with $\fsa$
the final state ones. In eqs.~(\ref{eq:spl})--(\ref{eq:sfA}) we have
shortened the notation, denoting e.g. $L=(\fsi,\pm)$ and $L=(\fsi,\fsa)$ 
simply by $\pm$ and $\fsa$ respectively, for any given particle label 
$\fsi$. This does not result in any ambiguities, since clearly the
equations above are formally identical for any $\fsi$.

The distribution function of parton $p$ in hadron $H_i$ is denoted by 
$f_p^{(H_i)}$, and the identities of initial-state partons 
$a$ and $b$ depend on $\mu$. In spite of
the fact that eqs.~(\ref{eq:spl}) and~(\ref{eq:smn}) are formally identical
to the analogous equations of ref.~\cite{Frixione:2005vw}, we stress that
the variables $z_\pm$ are different in the two cases, being (one of) 
the shower variables of \HWpp\ here, and of \HWs\ in 
ref.~\cite{Frixione:2005vw}. On the other hand, the variables
$\bx_{1i}$, $\bx_{2i}$, $\bx_{1f}$, and $\bx_{2f}$ (i.e. the
Bjorken $x$'s used by the PSMC) are the same for \HWpp\ as for \HWs. 
This statement is trivial in the context of standalone PSMC usage 
(given that the Bjorken $x$'s are just integration variables), 
but it is not when eqs.~(\ref{eq:spl})--(\ref{eq:sfA}) are used 
for the construction of the MC subtraction terms. In such
a case, in fact, the PSMC Bjorken $x$'s need be computed in terms
of those used in the NLO computation, and this is done in
\MCatNLO\ by means of the procedure called event projection,
described in details in ref.~\cite{Frixione:2002ik}.
Ultimately, event projection depends on the treatment of the
kinematics of the hard process by the PSMC. Since it is possible
to use the same procedure in \HWpp\ as in \HWs, it follows that
the variables above have the same analytic forms in the two cases.

The short-distance cross sections that appear in
eqs.~(\ref{eq:spl})--(\ref{eq:sfA}) read as follows:
\beqn
d\hat\sigma_{\mu}^{(\pm,l)}\xMCB &=& \frac{\as}{2\pi}\,
\frac{d\qtt_\pm}{\qtt_\pm}dz_\pm
P_{a^\prime b^\prime}(z_\pm)\,
\dsb_{\mu^\prime}^{(\pm,l)}\,\stepf_{\rm dead}^{(\pm,l)}\,,
\label{eq:shin}
\\
d\hat\sigma_{\mu}^{(\fsa,l)}\xMCB &=& \frac{\as}{2\pi}\,
\frac{d\qtt_{\fsa}}{\qtt_{\fsa}}dz_{\fsa}
P_{a^\prime b^\prime}(z_{\fsa},\qtt_{\fsa})\,
\dsb_{\mu^\prime}^{(\fsa,l)}\,\stepf_{\rm dead}^{(\fsa,l)}\,.
\label{eq:shout}
\eeqn
Their forms follow in a straightforward manner from considering how
any (collinear-based) PSMC deals with $1\to 2$ branchings.
They are constructed by multiplying the Born-level cross section
$\dsb_{\mu^\prime}^{(L,l)}$ by the relevant collinear or quasi-collinear
Altarelli-Parisi splitting kernels (here both denoted by $P$),
times the measure that appears in the Sudakov form factors.
This is PSMC-specific, and depends in particular on the shower
variables. Those of \HWpp\ are denoted by $z_{L}$
and $\qtt_{L}$, and are discussed in appendix~\ref{sec:HWppzka}.
Finally, the $\stepf$ functions in eqs.~(\ref{eq:shin}) 
and~(\ref{eq:shout}) are due to the fact that in general there
are phase-space regions where the PSMC cannot emit radiation
(called dead zones). The dead zones are PSMC-specific; those
relevant to \HWpp\ are given in appendix~\ref{sec:HWppzka}.
The identities of the partons involved in the branching, $a^\prime$
and $b^\prime$, and of the Born process, $\mu^\prime$, are fully
determined by $\mu$ and by $(L,l)$ -- their particular values
are irrelevant in what follows.

\section{Illustrative results\label{sec:results}}
In this section we present some illustrative results of the
implementation of \MCN\ for \HWpp, compared in each case with results
from the \HWs\ implementation. Both are available for the same
range of processes:  Higgs boson, single vector boson, vector boson
pair, heavy quark pair, single top (with and without associated $W$),
lepton pair and associated Higgs+$W/Z$ production in hadron collisions.
In most processes the results are very similar.  This is reassuring,
since the showering algorithms in the two event generators, and the
corresponding modified subtractions in the NLO calculations,
are quite different.  Where differences are seen, they can be ascribed
to changes in the parton showering algorithm,
particularly for heavy quarks, and to different modelling of
non-perturbative physics.  There is an overall
tendency for slightly more but softer gluon radiation in \HWpp.

For these comparisons \MCN\ was interfaced to the current versions
of the event generators, \HWpp\ v2.4.2 and \HW\ v6.520. 
All results are for the LHC at centre-of-mass energy 7 TeV, using 
the CTEQ6.6 NLO parton distributions~\cite{Nadolsky:2008zw},
and without the inclusion of the underlying event. Finally, we
have switched the intrinsic $\pt$ off in both \HWpp\ and \HWs.
 
\subsection{Higgs boson production}
\begin{figure}[htb]
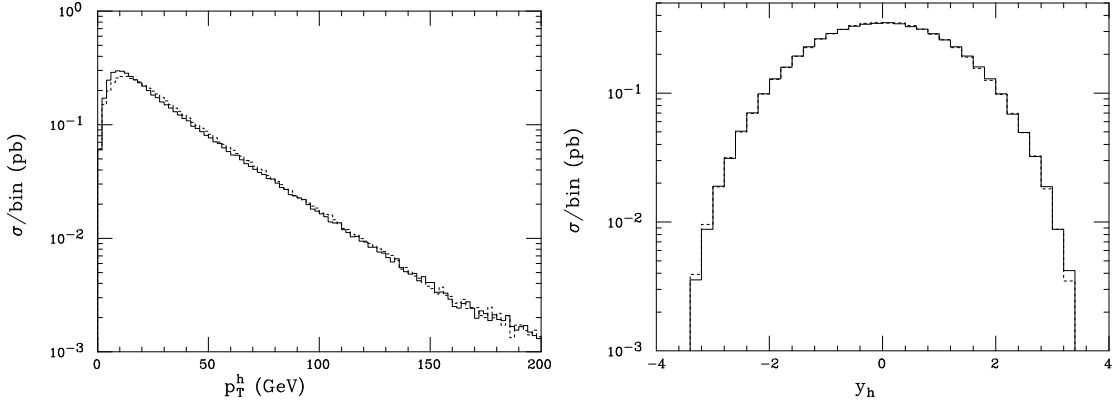

 \begin{center}
  \epsfig{figure=higgs_pt.ps,width=0.35\textwidth,angle=90}
  \epsfig{figure=higgs_y.ps,width=0.35\textwidth,angle=90}
\caption{\label{fig:higgs_pt} 
 \MCN\ results on Higgs boson production: Higgs transverse momentum
 (left)  and rapidity (right) distributions with \HWpp\ (solid) and 
\HWs\ (dashed).
}
 \end{center}
\end{figure}

In general the \MCN\ results on electroweak boson production using
\HWpp\ and \HWs\ show few significant differences.  Changes in the
boson transverse momentum distributions at low $\pt$ can be ascribed
to the softer QCD radiation in \HWpp. For example,
fig.~\ref{fig:higgs_pt} shows the transverse momentum and rapidity
distributions of a Standard-Model Higgs boson with a mass of 160 GeV.
Compared to \HWs, an increase is observed for $p_t<20$ GeV, due to the
slight softening of QCD radiation, but the rapidity distribution is
not affected.

\begin{figure}[htb]
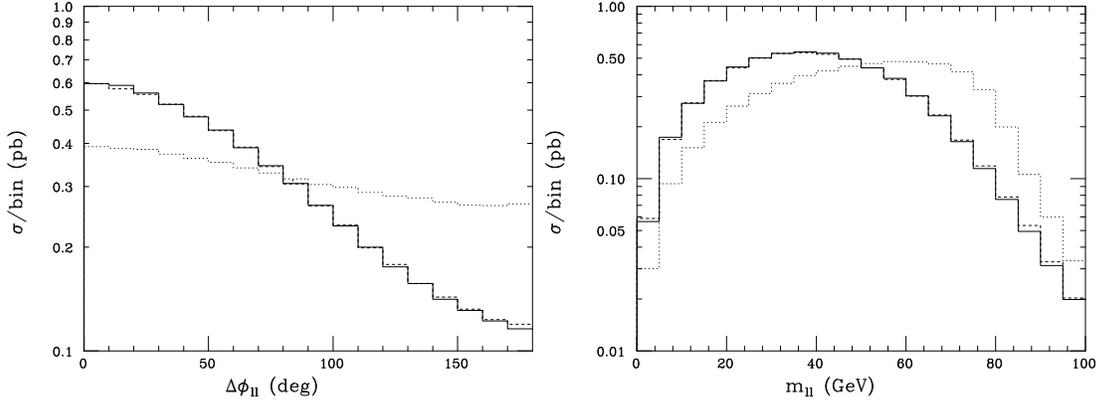

 \begin{center}
   \epsfig{figure=higgs_llaz.ps,width=0.35\textwidth,angle=90}
   \epsfig{figure=higgs_llm.ps,width=0.35\textwidth,angle=90}
\caption{\label{fig:higgs_ll} 
 \MCN\ results on Higgs $\to WW\to l\nu l\nu$ for \HWpp\ (solid) and
 for \HWs\ with (dashed) and without (dotted)  spin correlations.  Left panel:
 azimuthal angle between the charged leptons. Right
 panel: dilepton invariant mass.
}
 \end{center}
\end{figure}

Figure~\ref{fig:higgs_ll} shows the charged lepton correlations when
the Higgs boson decays to $WW\to l\nu l\nu$.  These distributions are not much
affected by soft QCD radiation and there is close agreement between the two
implementations.  However, clear differences are seen if earlier versions
of \HWs\ are used (up to v6.510, shown dotted), since spin correlations
between the $W$ bosons were not implemented before v6.520.

%%%%%%%%%%%%%%%%%%%%%%%%%%%%%%%%%%%%%%%%%%%%%%%%%%%%%%%%%%%%%%%%%%%
%%%%%%%%%%%%%%%%%%%%%%%%%%%%%%%%%%%%%%%%%%%%%%%%%%%%%%%%%%%%%%%%%%%

%%%%%%%%%%%%%%%%%%%%%%%%%%%%%%%%%%%%%%%%%%%%%%%%%%%%%%%%%%%%%%%%%%%
%%%%%%%%%%%%%%%%%%%%%%%%%%%%%%%%%%%%%%%%%%%%%%%%%%%%%%%%%%%%%%%%%%%

In refs.~\cite{Anastasiou:2008ik,Anastasiou:2009bt} the effects of
acceptance cuts on searches for a Higgs boson in the $WW\to l\nu l\nu$
decay channel were studied. The cuts were chosen to mimic realistic
event selection as applied in experimental searches for Higgs bosons in
this channel:
\begin{enumerate}
\item the leading charged lepton should be within $|\eta|<2$ and
  $30\,\mathrm{GeV}<\pt<55\,\mathrm{GeV}$;
\item the trailing charged lepton has to fulfill  $|\eta|<2$ and
  $\pt>25\,\mathrm{GeV}$;
\item the invariant mass of the charged lepton pair is restricted to
  $12\,\mathrm{GeV}<m_{ll}<40\,\mathrm{GeV}$;
\item the missing transverse energy has to exceed $50\,\mathrm{GeV}$;
\item the opening angle between the two charged leptons in the
  transverse plane has to be smaller then $\phi_{ll}<\pi/4$;
\item there should be no jet with $\pt>25\,\mathrm{GeV}$ and
  $|\eta|<2.5$ in the event (jet-veto).
\end{enumerate}

Higher-order QCD corrections were seen to
have significant impact on the efficiency of these cuts, and \MCN\
with \HWs\ was found to give good agreement with NNLO
calculations of acceptance effects.  In table~\ref{tab:higgs_accept}
we show the corresponding results for  \MCN\ with \HWpp.
As can be seen, the efficiencies predicted by the two event generators
are similar, though not identical. The
cuts applied can be divided into two sets: cuts on kinematic variables of the
final state leptons (such as invariant mass, transverse momentum, rapidity), 
and cuts on the hadronic structure of the event (jet-veto). While the
acceptances of the leptonic cuts are very similar between \HWpp\ and
\HWs\ (see also figure~\ref{fig:higgs_ll}), the difference in the overall 
efficiency arises from the different hadronic structure in the low momentum 
region (see also figure~\ref{fig:higgs_pt}). This demonstrates again how
the uncertainty on a jet-veto efficiency has to be studied very
carefully when applied in experimental searches.

%%%%%%%%%%%%%%%%%%%%%%%%%%%%%%%%%%%%%%%%%%%%%%%%%%%%%%%%%%%%%%%%%%%
\begin{table}[htb]
 \begin{center}
\begin{tabular}{|l|c|c|c|}
\hline
Generator & $\sigma_\mathrm{inc}$ [fb] & $\sigma_\mathrm{acc}$ [fb] &
$\varepsilon$ [\%] \\
\hline\hline
\HWpp & 5.76 & 0.479 & $8.32\,\pm\,0.04$ \\
\HWs  & 5.76 & 0.444 & $7.71\,\pm\,0.04$ \\
\hline
\end{tabular}

\caption{\label{tab:higgs_accept} 
 \MCN\ results on the acceptance for Higgs $\to WW\to l\nu l\nu$ after
 the cuts indicated in the text. The errors on the efficiencies are the
 statistical uncertainties.
}
 \end{center}
\end{table}
%%%%%%%%%%%%%%%%%%%%%%%%%%%%%%%%%%%%%%%%%%%%%%%%%%%%%%%%%%%%%%%%%%%

\subsection{Top quark pair production}

The \HWs\ and \HWpp\ event generators differ significantly in their
treatment of heavy quarks.  The main differences are
\begin{enumerate}
\item The kinematics of the parton shower.   The energy-angle shower
  variables used in \HWs\ led to a sharp angular cutoff at $\theta\sim
  m/E$ in the radiation pattern from a quark of mass $m$ and energy
  $E$ (the so-called {\it dead cone}), whereas \HWpp\ uses a more
  covariant formulation that allows emission at lower 
  angles~\cite{Gieseke:2003rz}.
\item The introduction of mass corrections to the parton splitting
  functions in \HWpp,  following the `quasi-collinear' prescription of
  ref.~\cite{Catani:2000ef}.
\item An improved treatment of QCD radiation in top decay in \HWpp,
  developed in refs.~\cite{Gieseke:2003rz,Hamilton:2006ms}, which
  ensures a  better angular distribution and removes the need for an
  {\it ad hoc}   infrared cutoff.
\end{enumerate}

As a result of these improvements, differences in the results of
\MCN\ with \HWs\ and \HWpp\ for processes involving heavy quarks are
somewhat more pronounced than those in electroweak boson
production. The most significant effect is a softer spectrum
of QCD radiation in top quark production and decay in \HWpp.

\begin{figure}[htb]
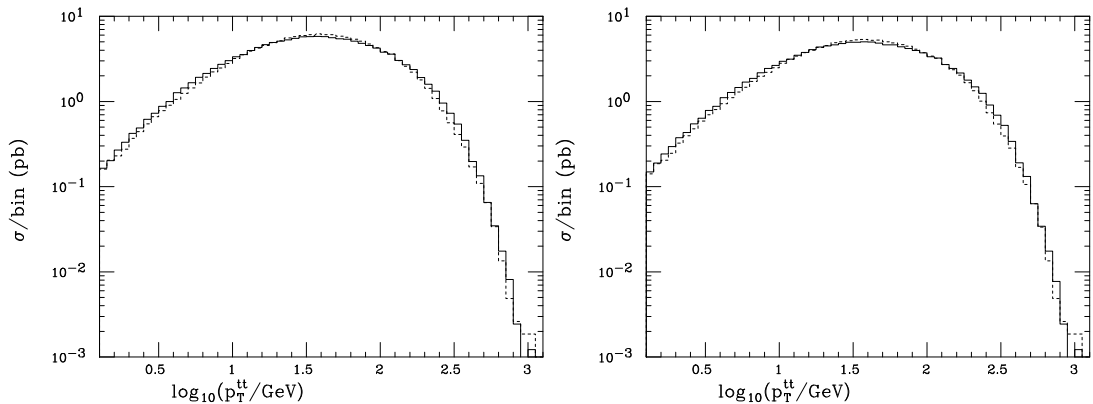

 \begin{center}
   \epsfig{figure=tt_ttlogpt.ps,width=0.35\textwidth,angle=90}
   \epsfig{figure=tt_ttlogpt_cuts.ps,width=0.35\textwidth,angle=90}
\caption{\label{fig:tt_ttpt} 
 \MCN\ results on top quark pair production: $\log_{10}$ of the $t\bar t$ 
 transverse momentum distribution with \HWpp\ (solid) and \HWs\ (dashed).
 Left/right panel: without/with acceptance cuts.
}
 \end{center}
\end{figure}

\begin{figure}[htb]
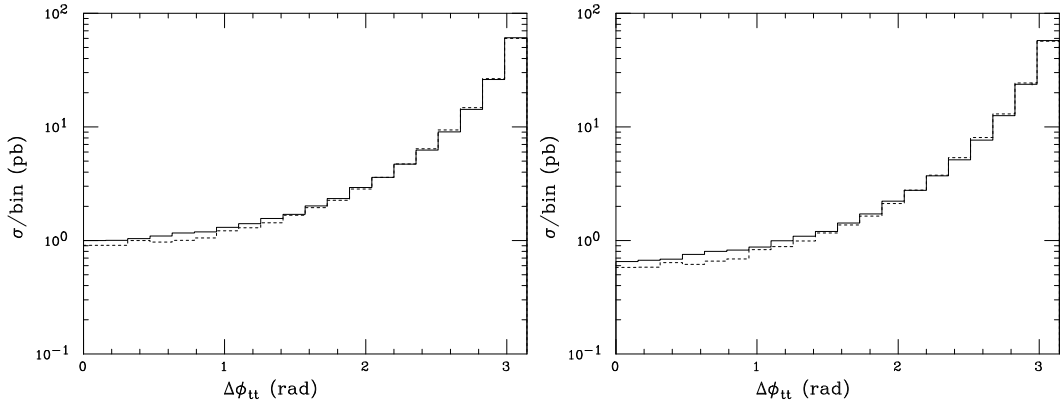

\begin{center}
 \epsfig{figure=tt_ttaz.ps,width=0.35\textwidth,angle=90}
 \epsfig{figure=tt_ttaz_cuts.ps,width=0.35\textwidth,angle=90}
\caption{\label{fig:tt_ttaz} 
\MCN\ results on top quark pair production: $t\bar t$ azimuthal separation
with \HWpp\ (solid) and \HWs\ (dashed).
Left/right panel: without/with acceptance cuts.
}
\end{center}
\end{figure}

\begin{figure}[htb]
 \begin{center}
   \epsfig{figure=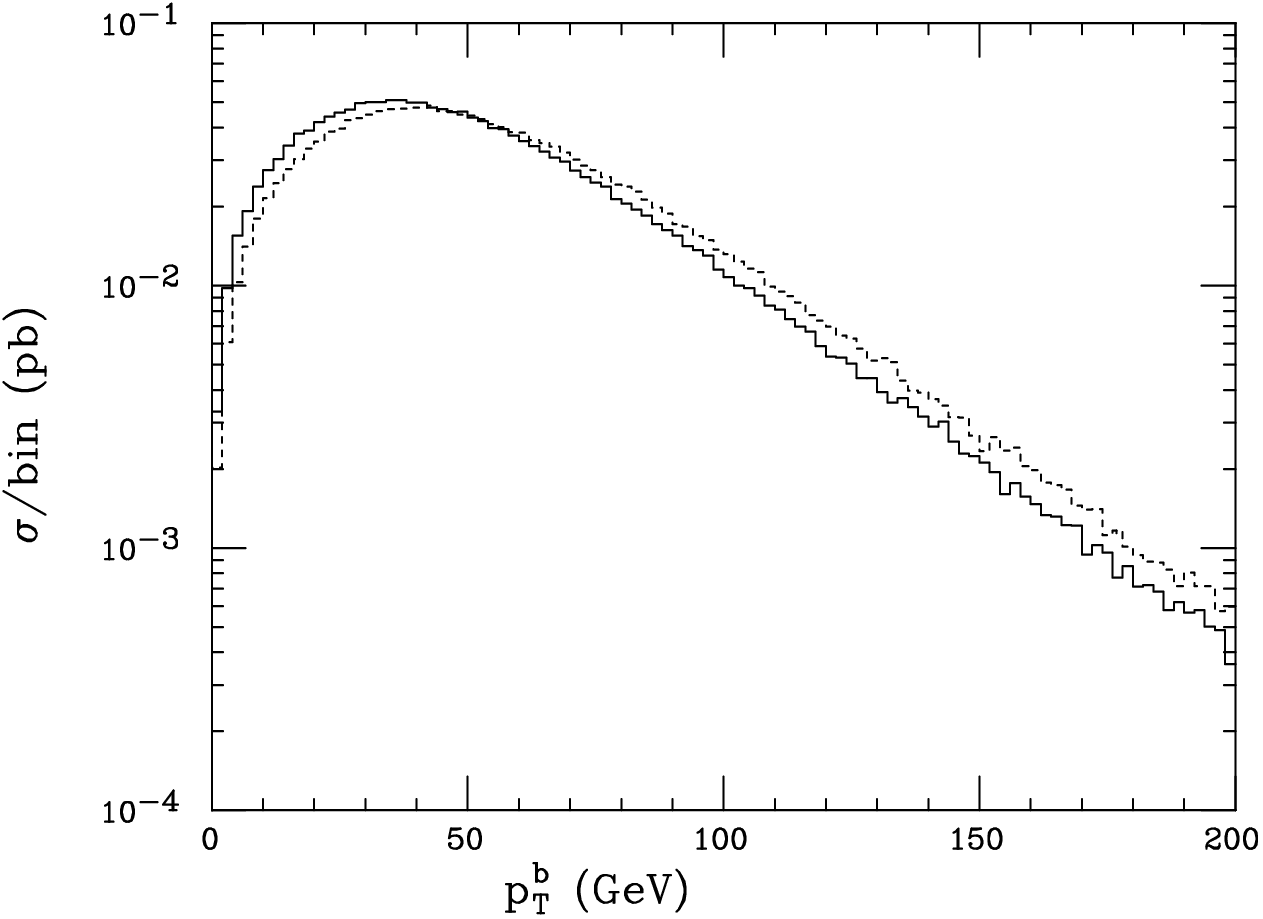,width=0.35\textwidth,angle=90}
   \epsfig{figure=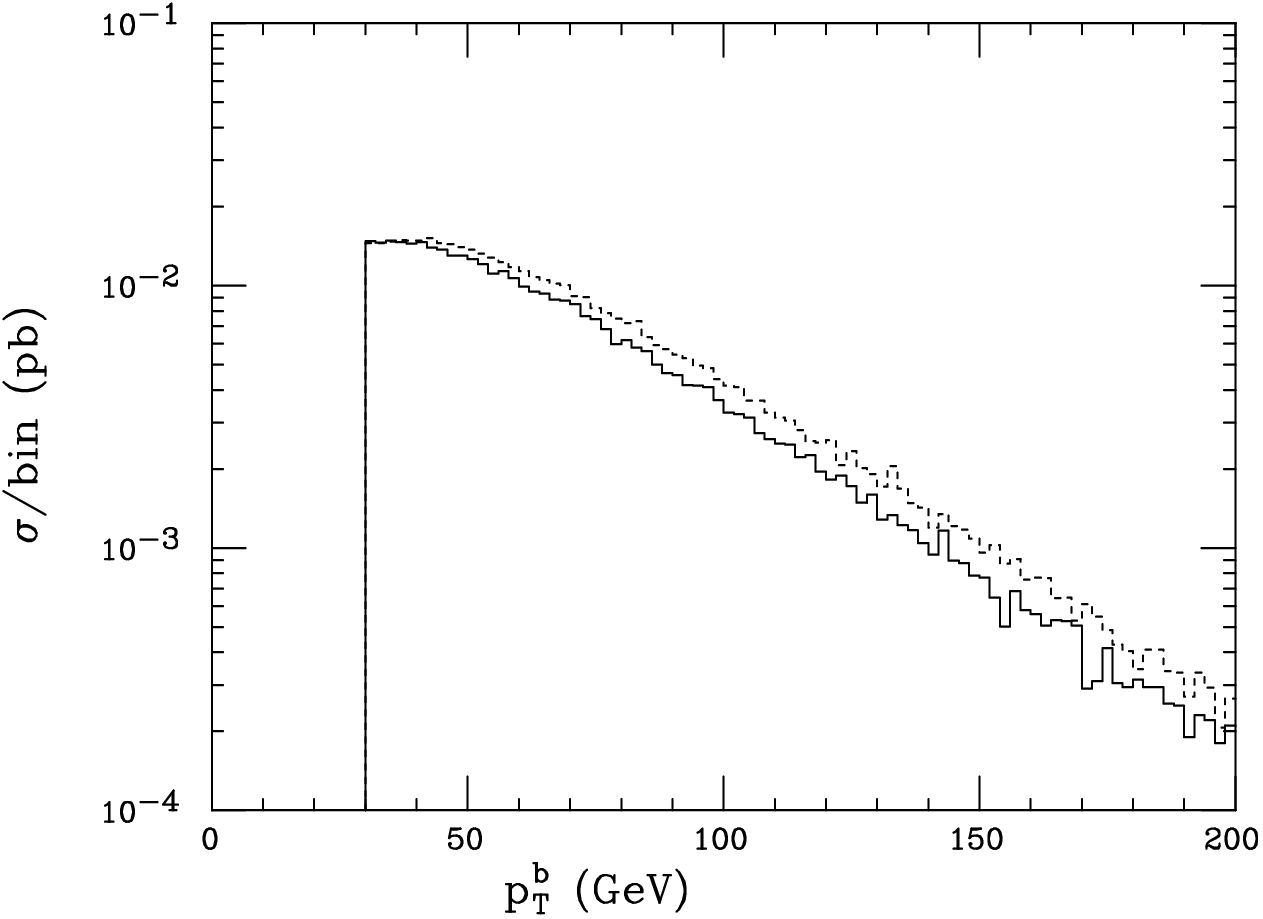,width=0.35\textwidth,angle=90}
\caption{\label{fig:tt_bpt} 
 \MCN\ results on top quark pair production: $b$-quark transverse momentum 
 distribution with \HWpp\ (solid) and \HWs\ (dashed).
 Left/right panel: without/with acceptance cuts.
}
 \end{center}
\end{figure}

\begin{figure}[htb]
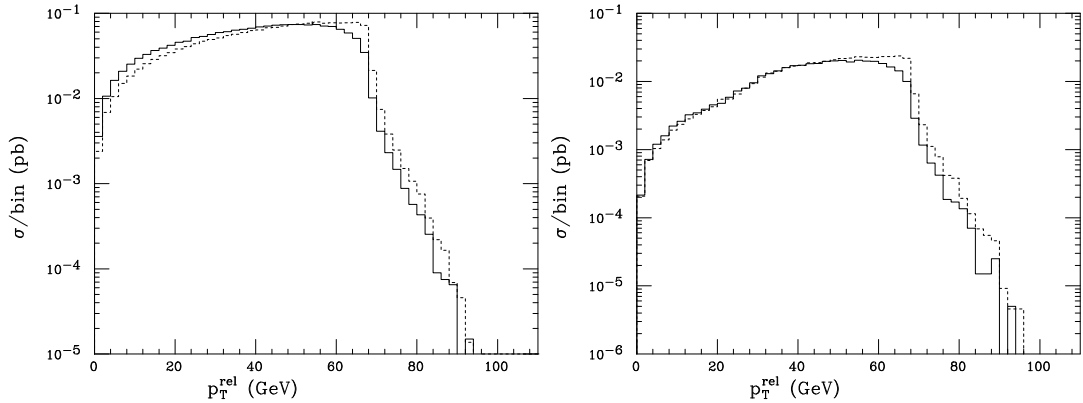

 \begin{center}
   \epsfig{figure=tt_bptrel.ps,width=0.35\textwidth,angle=90}
   \epsfig{figure=tt_bptrel_cuts.ps,width=0.35\textwidth,angle=90}
\caption{\label{fig:tt_bptrel} 
 \MCN\ results on top quark pair production: $b$-quark transverse 
 momentum (relative to the direction of flight of the top) distribution 
 with \HWpp\ (solid) and \HWs\ (dashed).
 Left/right panel: without/with acceptance cuts.
}
 \end{center}
\end{figure}

\begin{figure}[htb]
 \begin{center}
   \epsfig{figure=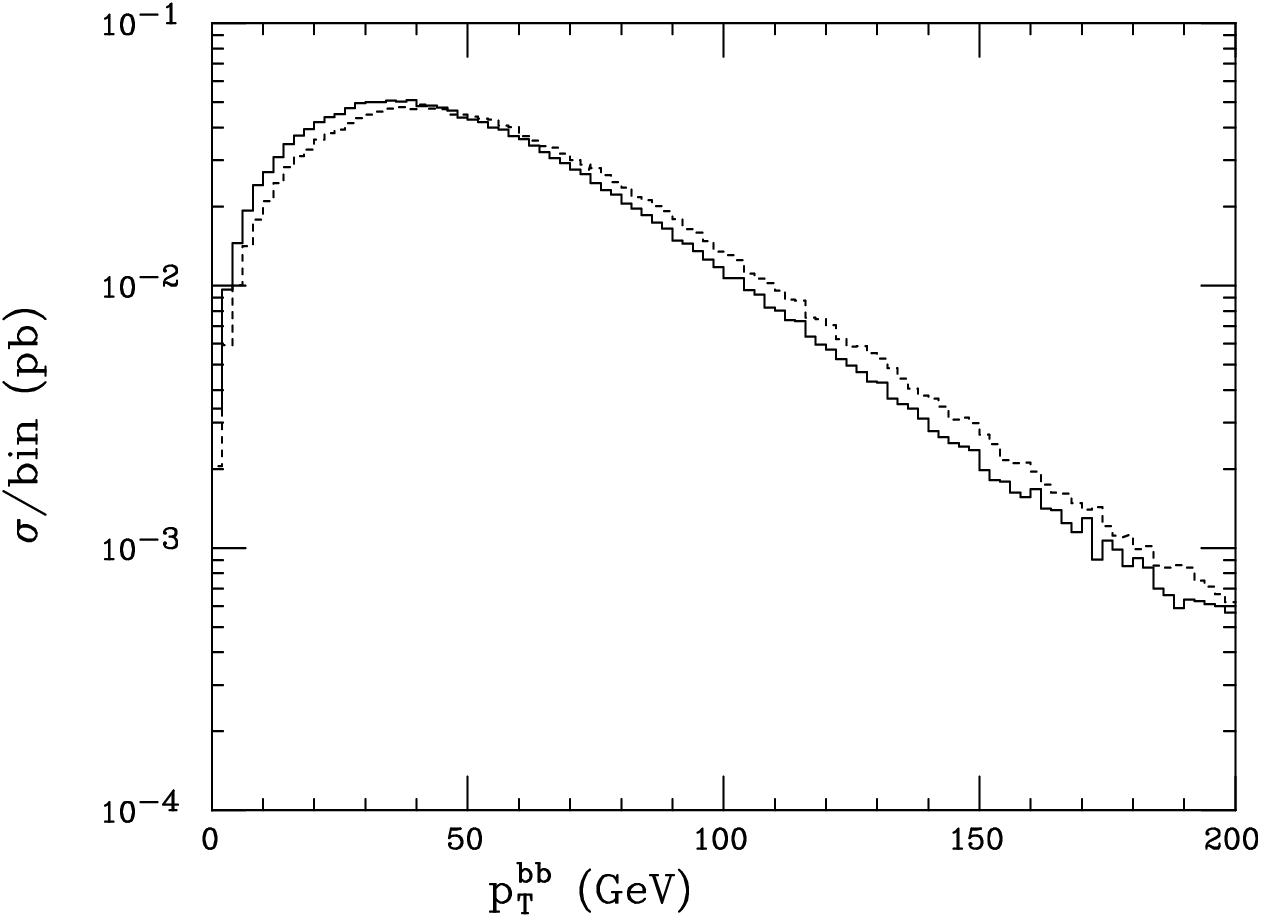,width=0.35\textwidth,angle=90}
   \epsfig{figure=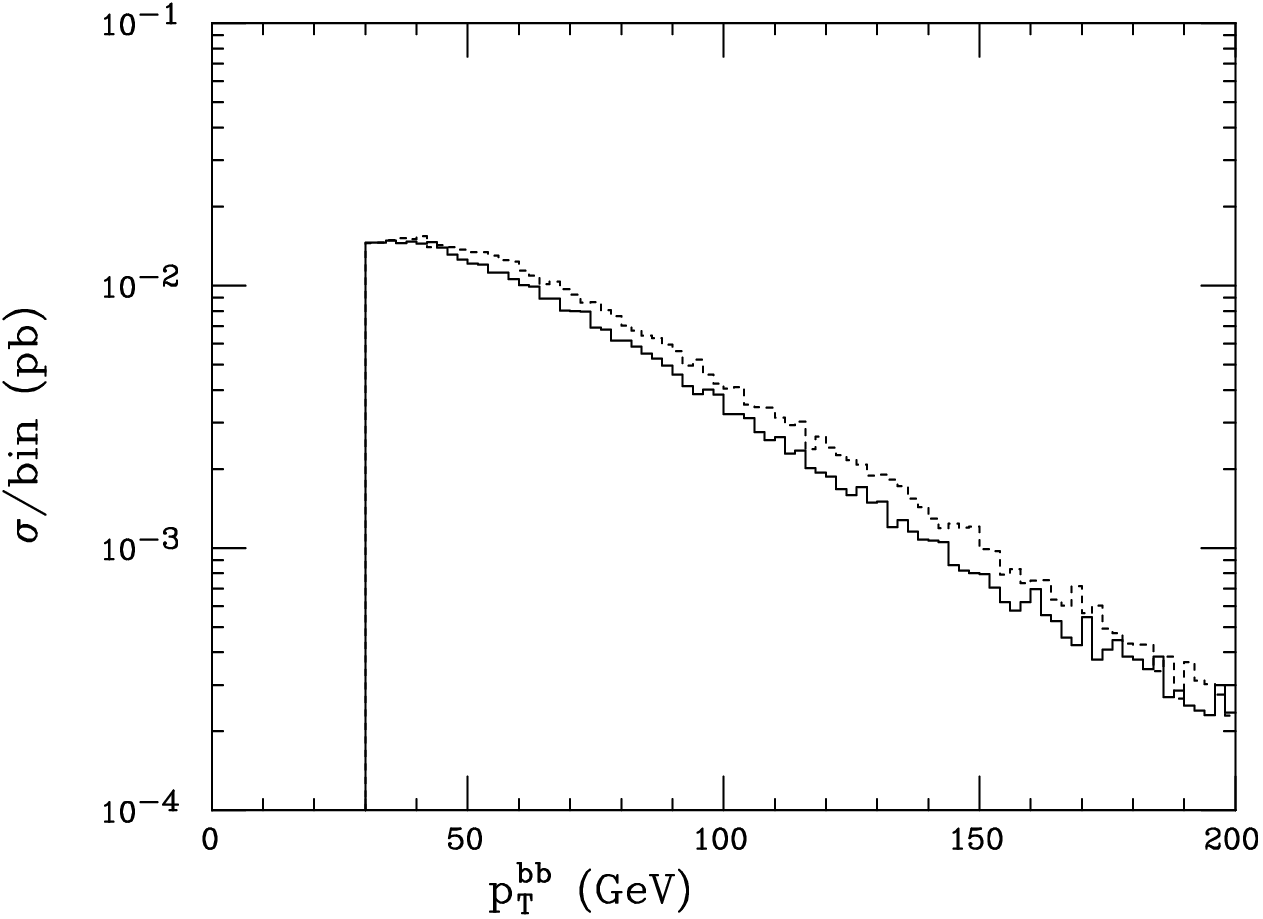,width=0.35\textwidth,angle=90}
\caption{\label{fig:tt_bbpt} 
 \MCN\ results on top quark pair production: $b\bar b$ transverse momentum 
 distribution with \HWpp\ (solid) and \HWs\ (dashed).
 Left/right panel: without/with acceptance cuts.
}
 \end{center}
\end{figure}

\begin{figure}[htb]
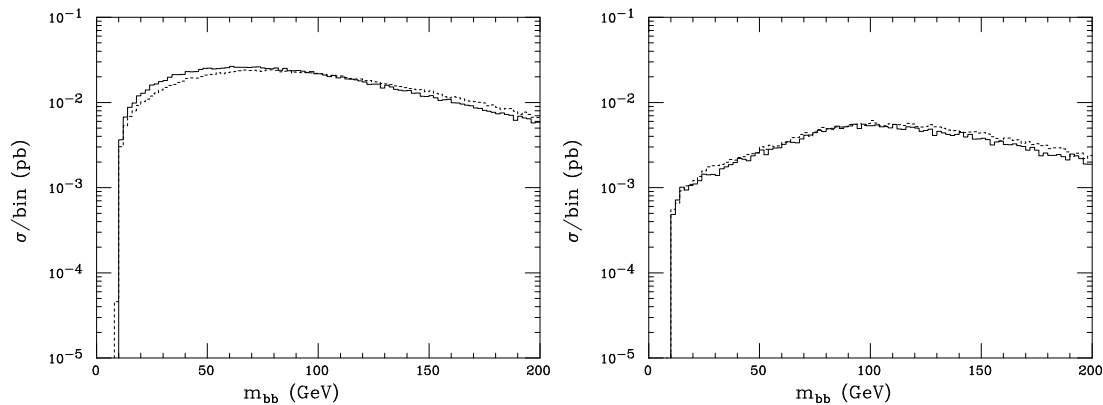

 \begin{center}
   \epsfig{figure=tt_bbm.ps,width=0.35\textwidth,angle=90}
   \epsfig{figure=tt_bbm_cuts.ps,width=0.35\textwidth,angle=90}
\caption{\label{fig:tt_bbm} 
 \MCN\ results on top quark pair production: $b\bar b$ invariant mass 
 distribution with \HWpp\ (solid) and \HWs\ (dashed).
 Left/right panel: without/with acceptance cuts.
}
 \end{center}
\end{figure}

Figures~\ref{fig:tt_ttpt} and~\ref{fig:tt_ttaz} show the resulting 
effects on the $t\bar t$
transverse momentum spectrum and azimuthal separation, respectively.
In these and subsequent figures we also show the distributions after
acceptance cuts. For top quark pair production, these are defined as follows:
in figs.~\ref{fig:tt_ttpt} and~\ref{fig:tt_ttaz}, we require the transverse
momenta of the $t$ and $\bar{t}$ to be larger than $30$~GeV, and the 
absolute values of their rapidities to be smaller than 2.5. On the
other hand, in figs.~\ref{fig:tt_bpt}--\ref{fig:tt_bbm}, which were
obtained by letting the top quarks decay leptonically, we require the
transverse momenta (absolute values of the rapidities) of the ``visible'' 
decay products, i.e. the $b$ and $\bar{b}$ quarks and the charged 
leptons $l^\pm$, to be larger than $30$~GeV (smaller than 2.5).

Figures~\ref{fig:tt_bpt} and \ref{fig:tt_bptrel} show aspects of
the distribution of $b$ quarks from decay of the top pair: the
transverse momentum relative to the beam direction in
fig.~\ref{fig:tt_bpt} and relative to the  direction of motion
of the parent top in fig.~\ref{fig:tt_bptrel}.  Decay angular
correlations due to the top polarization are included following the
prescription of ref.~\cite{Frixione:2007zp}.  Here again there is
some softening and smearing of the distribution in \HWpp.
A corresponding softening of the transverse momentum and invariant
mass distributions of the pair of $b$ quarks may be seen in 
figs.~\ref{fig:tt_bbpt} and \ref{fig:tt_bbm}.

%%%%%%%%%%%%%%%%%%%%%%%%%%%%%%%%%%%%%%%%%%%%%%%%%%%%%%%%%%%%%%%%%%%
%%%%%%%%%%%%%%%%%%%%%%%%%%%%%%%%%%%%%%%%%%%%%%%%%%%%%%%%%%%%%%%%%%%

%%%%%%%%%%%%%%%%%%%%%%%%%%%%%%%%%%%%%%%%%%%%%%%%%%%%%%%%%%%%%%%%%%%
%%%%%%%%%%%%%%%%%%%%%%%%%%%%%%%%%%%%%%%%%%%%%%%%%%%%%%%%%%%%%%%%%%%

%%%%%%%%%%%%%%%%%%%%%%%%%%%%%%%%%%%%%%%%%%%%%%%%%%%%%%%%%%%%%%%%%%%
%%%%%%%%%%%%%%%%%%%%%%%%%%%%%%%%%%%%%%%%%%%%%%%%%%%%%%%%%%%%%%%%%%%

%%%%%%%%%%%%%%%%%%%%%%%%%%%%%%%%%%%%%%%%%%%%%%%%%%%%%%%%%%%%%%%%%%%
%%%%%%%%%%%%%%%%%%%%%%%%%%%%%%%%%%%%%%%%%%%%%%%%%%%%%%%%%%%%%%%%%%%

%%%%%%%%%%%%%%%%%%%%%%%%%%%%%%%%%%%%%%%%%%%%%%%%%%%%%%%%%%%%%%%%%%%
%%%%%%%%%%%%%%%%%%%%%%%%%%%%%%%%%%%%%%%%%%%%%%%%%%%%%%%%%%%%%%%%%%%

%%%%%%%%%%%%%%%%%%%%%%%%%%%%%%%%%%%%%%%%%%%%%%%%%%%%%%%%%%%%%%%%%%
%%%%%%%%%%%%%%%%%%%%%%%%%%%%%%%%%%%%%%%%%%%%%%%%%%%%%%%%%%%%%%%%%%%

\subsection{Single top production}

\begin{figure}[htb]
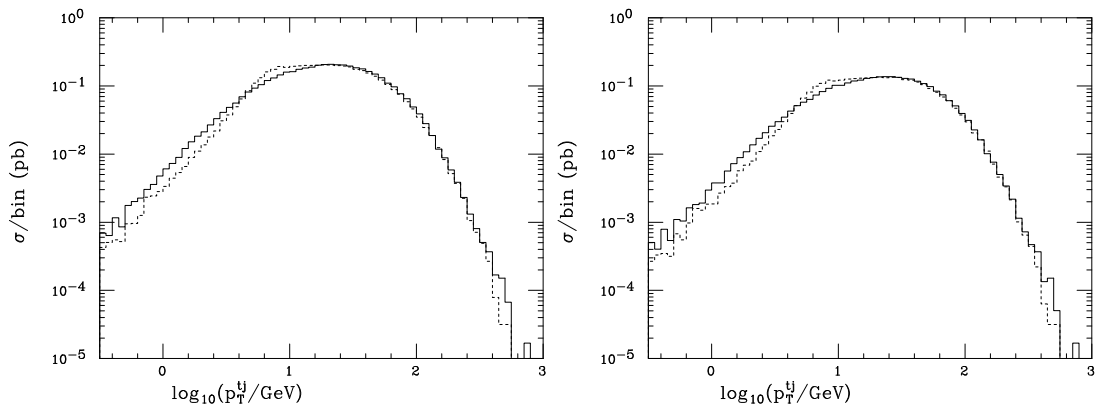

 \begin{center}
   \epsfig{figure=st_tj1pt.ps,width=0.35\textwidth,angle=90}
   \epsfig{figure=st_tj1pt_cuts.ps,width=0.35\textwidth,angle=90}
\caption{\label{fig:st_tj1pt} 
 \MCN\ results on single top production: $\log_{10}$ of the transverse 
 momentum distribution of the top-hardest jet pair, with \HWpp\ (solid) 
 and \HWs\ (dashed). Left/right panel: without/with acceptance cuts.
}
 \end{center}
\end{figure}

Comparisons between results on single top production\footnote{We have
limited ourselves to considering here top quark production (i.e., antitop
quark production is not included) in the $t$ channel.} from \MCN\ with
\HWpp\ and \HWs\ reveal similar basic features to those in top quark
pair production.  There is a general softening of distributions due to
softer QCD radiation in \HWpp, illustrated here by the top plus leading
jet transverse momentum distribution, fig.~\ref{fig:st_tj1pt}.
The jets have been defined with the $\kt$ algorithm as implemented
by FastJet~\cite{Cacciari:2005hq}, with $R=0.5$ and by requiring each jet
to have transverse momentum larger than 10~GeV. Furthermore, we have
eliminated from the list of our jets the one that contains the $b$-flavoured
hadron that emerges from the decay of the top. Apart from the jet
cuts, when imposing acceptance cuts (except in the case of 
fig.~\ref{fig:st_cos}, see below) we have required the top to have 
transverse momentum larger than 20~GeV, and the absolute value of its 
rapidity to be smaller than 2.5.

\begin{figure}[htb]
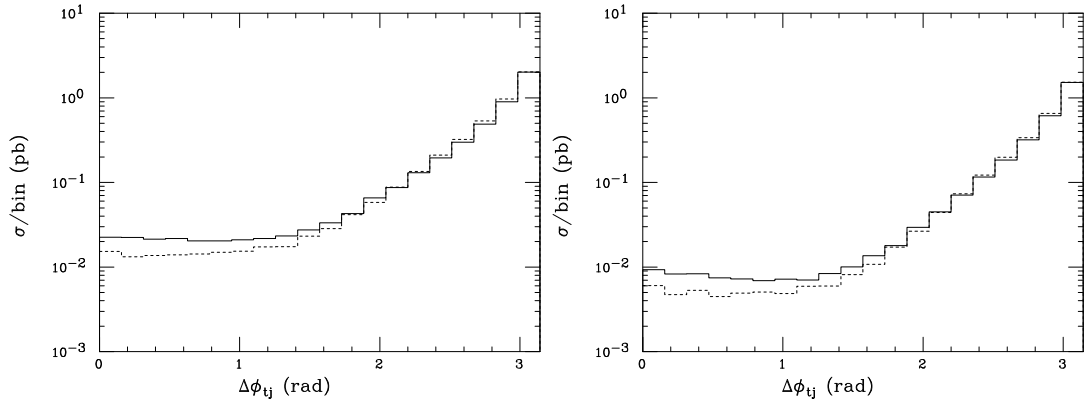

 \begin{center}
   \epsfig{figure=st_tj1az.ps,width=0.35\textwidth,angle=90}
   \epsfig{figure=st_tj1az_cuts.ps,width=0.35\textwidth,angle=90}
\caption{\label{fig:st_tj1az} 
 \MCN\ results on single top production: relative azimuth distribution
 between the top and the hardest jet, with \HWpp\ (solid) and \HWs\ (dashed).
 Left/right panel: without/with acceptance cuts.
}
 \end{center}
\end{figure}

The distribution of the top--leading jet relative azimuth
(fig.~\ref{fig:st_tj1az}) shows an increase in same-side emission,
while the angular separation (measured in the rest frame of the top)
between the charged lepton in leptonic top decays and the hardest 
non-$b$ jet (fig.~\ref{fig:st_cos}) tends to be slightly larger.
In the right panel of fig.~\ref{fig:st_cos}, we have imposed the 
same acceptance cuts as in ref.~\cite{Frixione:2007zp}; namely,
$\pt(b)>20$~GeV, $\abs{\eta(b)}<2$, $\pt(l)>10$~GeV, $\abs{\eta(l)}<2.5$, 
$\pt(\nu)>20$~GeV, $\pt(j)>20$~GeV, and $\abs{\eta(j)}<2.5$. 

\begin{figure}[htb]
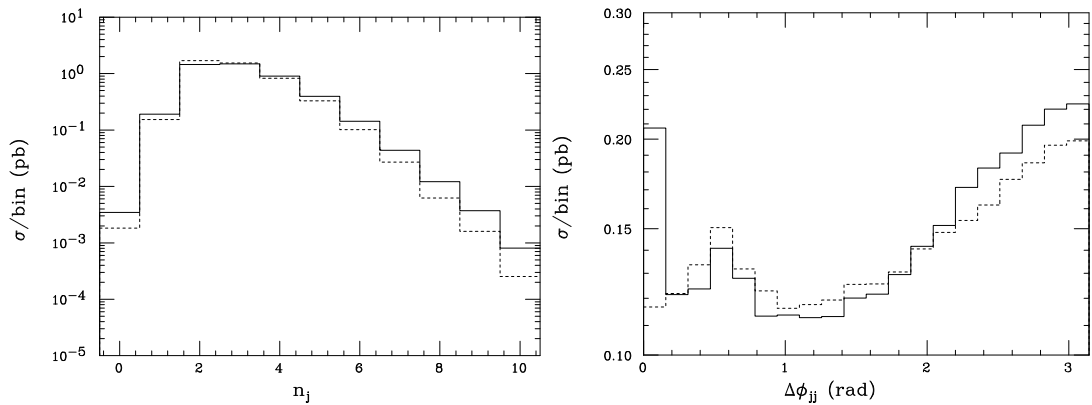

 \begin{center}
   \epsfig{figure=st_njets.ps,width=0.35\textwidth,angle=90}
   \epsfig{figure=st_j1j2az.ps,width=0.35\textwidth,angle=90}
\caption{\label{fig:st_j1j2az} 
 \MCN\ results on single top production:  number of jets (left panel)
 and azimuthal separation between the two hardest jets (right panel),
 with \HWpp\ (solid) and \HWs\ (dashed).
}
 \end{center}
\end{figure}

In accordance with the general tendency for more radiation in
\HWpp, the overall jet activity (fig.~\ref{fig:st_j1j2az}, left
panel) is somewhat higher than in \HWs.
The relative azimuthal distributions of the leading and
next-to-leading jets (fig.~\ref{fig:st_j1j2az}, right panel) are similar 
apart from  an unexpected feature in the \HWpp\ distribution at low values.
The corresponding jets have different rapidities but are aligned in
azimuth.  This appears to be a feature of the non-perturbative
splitting of high-mass clusters connected to the beam remnant in \HWpp.

%%%%%%%%%%%%%%%%%%%%%%%%%%%%%%%%%%%%%%%%%%%%%%%%%%%%%%%%%%%%%%%%%%%
%%%%%%%%%%%%%%%%%%%%%%%%%%%%%%%%%%%%%%%%%%%%%%%%%%%%%%%%%%%%%%%%%%%

%%%%%%%%%%%%%%%%%%%%%%%%%%%%%%%%%%%%%%%%%%%%%%%%%%%%%%%%%%%%%%%%%%%
%%%%%%%%%%%%%%%%%%%%%%%%%%%%%%%%%%%%%%%%%%%%%%%%%%%%%%%%%%%%%%%%%%%

%%%%%%%%%%%%%%%%%%%%%%%%%%%%%%%%%%%%%%%%%%%%%%%%%%%%%%%%%%%%%%%%%%%
\begin{figure}[htb]
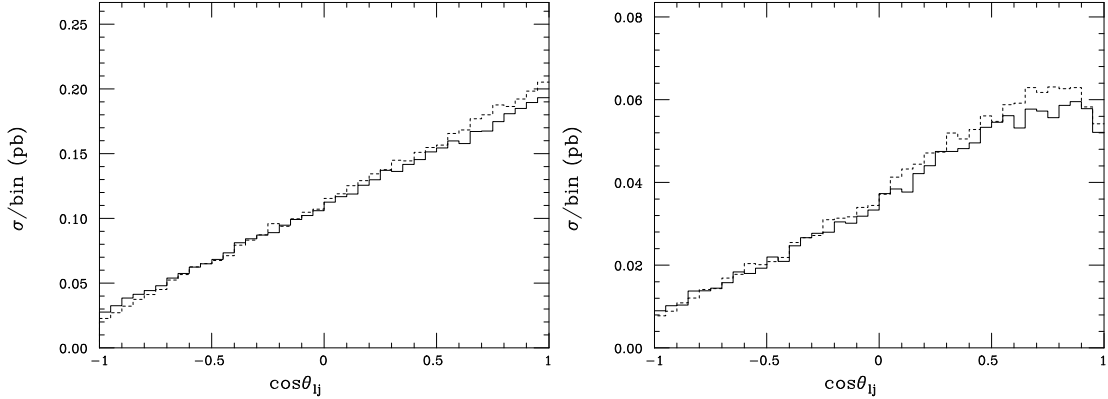

\begin{center}
  \epsfig{figure=st_cos.ps,width=0.35\textwidth,angle=90}
  \epsfig{figure=st_cos_cuts.ps,width=0.35\textwidth,angle=90}
\caption{\label{fig:st_cos} 
\MCN\ results on single top production: angle (in the top rest frame)
between lepton from top decay and hardest non-$b$ jet with \HWpp\ (solid) 
and \HWs\ (dashed). Left/right panel: without/with acceptance cuts.
}
\end{center}
\end{figure}
%%%%%%%%%%%%%%%%%%%%%%%%%%%%%%%%%%%%%%%%%%%%%%%%%%%%%%%%%%%%%%%%%%%

%%%%%%%%%%%%%%%%%%%%%%%%%%%%%%%%%%%%%%%%%%%%%%%%%%%%%%%%%%%%%%%%%%%
%%%%%%%%%%%%%%%%%%%%%%%%%%%%%%%%%%%%%%%%%%%%%%%%%%%%%%%%%%%%%%%%%%%

\begin{figure}[htb]
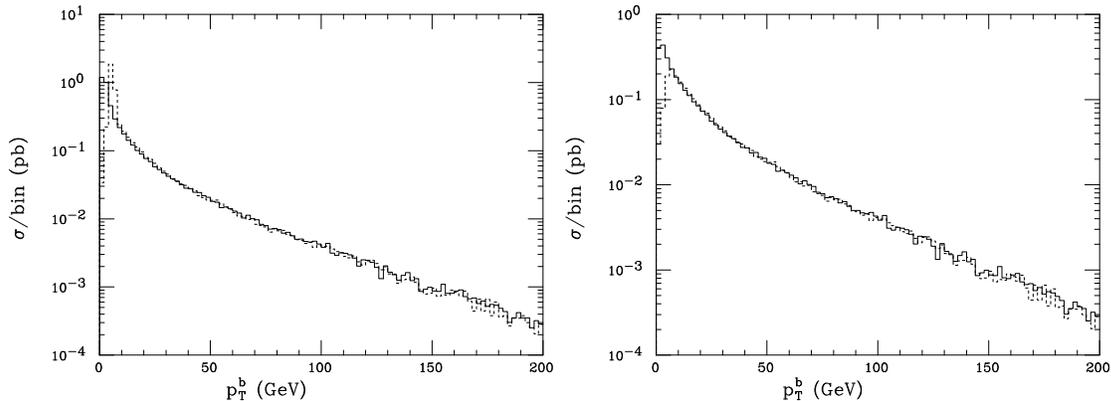

 \begin{center}
   \epsfig{figure=st_b2pt.ps,width=0.35\textwidth,angle=90}
   \epsfig{figure=st_b2pt_cuts.ps,width=0.35\textwidth,angle=90}
\caption{\label{fig:st_b2pt} 
 \MCN\ results on single top production: transverse momentum
 distributions of $b$-hadrons not from top decay with \HWpp\ (solid)
 and \HWs\ (dashed). Left/right panels: without/with a cut on rapidity $|y|<3$.
}
 \end{center}
\end{figure}

\begin{figure}[htb]
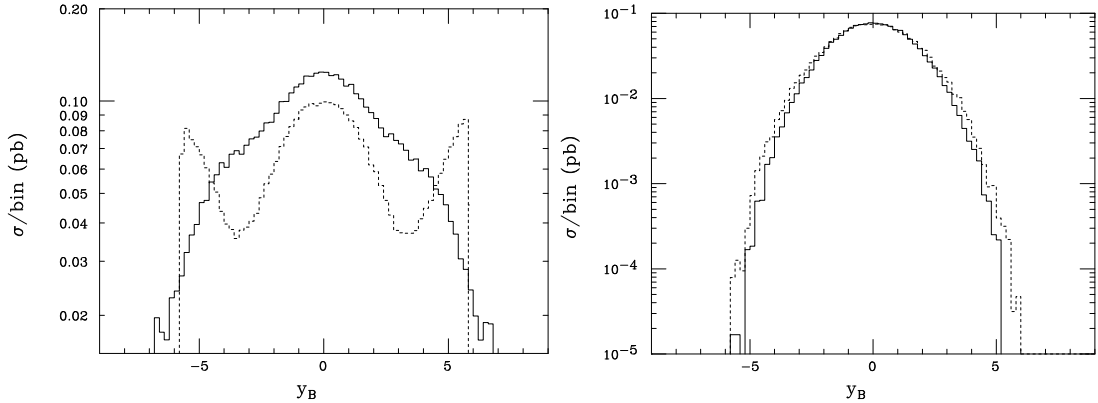

 \begin{center}
   \epsfig{figure=st_b2y.ps,width=0.35\textwidth,angle=90}
   \epsfig{figure=st_b2y_cuts.ps,width=0.35\textwidth,angle=90}
\caption{\label{fig:st_b2y} 
 \MCN\ results on single top production: rapidity distributions of 
 $b$-hadrons not from top decay with \HWpp\ (solid) and \HWs\ (dashed).  
 Left/right panels: without/with a cut on transverse momentum $\pt>10$ GeV.
}
 \end{center}
\end{figure}

In other respects the treatment of non-perturbative effects looks more
physical in \HWpp.  Figures~\ref{fig:st_b2pt} and~\ref{fig:st_b2y} 
show the transverse
momentum and rapidity distributions of $b$-hadrons not from top decay,
which come mainly from parton showering of initial-state $b$ quarks.
In \HWs\ these distributions have pathologies at low $\pt$ and high
rapidity, arising from its simplified treatment of heavy quark
showering and the model of non-perturbative $g\to b\bar b$ splitting
used in matching the shower to the beam hadron.  The latter gives rise
to a deficit at $\pt<m_b$ and peaks around $|y|\sim 5$, which are
less prominent in the model used in \HWpp, which has a smoother transition
to the non-perturbative regime.  As shown in the right-hand panels of
figs.~\ref{fig:st_b2pt} and~\ref{fig:st_b2y},  a cut on $y$ or $\pt$ removes 
most of the model dependence and yields much closer agreement between 
results from the two generators. In particular, in the kinematic
region in which the $b$-hadrons are observable, the differences
between \HWpp\ and \HWs\ are small. This also implies that, owing to
the fact that the inclusive $b$-hadron cross section predicted by the
two PSMCs is identical (up to events with multiple $b$-hadrons, arising
from $g\to b\bar{b}$ branchings in the shower), the impact of a veto cut 
will be similar in \HWpp\ and in \HWs.

%%%%%%%%%%%%%%%%%%%%%%%%%%%%%%%%%%%%%%%%%%%%%%%%%%%%%%%%%%%%%%%%%%%
%%%%%%%%%%%%%%%%%%%%%%%%%%%%%%%%%%%%%%%%%%%%%%%%%%%%%%%%%%%%%%%%%%%

%%%%%%%%%%%%%%%%%%%%%%%%%%%%%%%%%%%%%%%%%%%%%%%%%%%%%%%%%%%%%%%%%%%
%%%%%%%%%%%%%%%%%%%%%%%%%%%%%%%%%%%%%%%%%%%%%%%%%%%%%%%%%%%%%%%%%%%

We conclude this section by mentioning that we have also considered
single-top production in association with a $W$ boson, and compared
\HWpp\ with \HWs\ results for both the DR and DS definitions of the
$Wt$ cross section (see ref.~\cite{Frixione:2008yi}). We did not
attempt a full phenomenological study of this process, but limited
ourselves to considering the impact of a $\pt$-veto imposed on
the second-hardest $b$-hadron, which as discussed in 
ref.~\cite{Frixione:2008yi} is a rather effective way of reducing
the $Wt$-$t\bar{t}$ interference. We have found that \HWpp\ follows
the same pattern as \HWs. This is reassuring, since it implies that
the definition of $Wt$ production as a separate process at the NLO
is independent of the PSMC used in the simulations.

\section{Conclusions\label{sec:concl}}
In this paper we have presented the calculations necessary to match
the Parton Shower Monte Carlo \HWpp\ with any NLO QCD computation
in the context of the \MCatNLO\ formalism. The matching has then 
been achieved in practice for all processes which were already
interfaced to Fortran \HW. A few selected \MCN/\HWpp\ results
obtained in this way have also been shown here, and compared
to those obtained with \MCN/\HW.

From the technical point of view, the calculations performed here
are of a complexity comparable with those reported in 
refs.~\cite{Frixione:2002ik,Frixione:2003ei,Frixione:2005vw},
which were relevant to Fortran \HW. The corresponding computer 
programmes also behave in a fairly similar way. As expected, in
those phase-space regions dominated by hard emissions, the results
of \MCN/\HWpp\ and \MCN/\HWs\ coincide. On the other hand,
differences (usually small) can be seen where multiple-parton emission 
plays a dominant role, with \HWpp\ typically giving a larger number of
partons than \HW, but with smaller energies. In a very few cases 
larger discrepancies can be seen, and we have commented them in 
the text.

The results presented here provide all the ingredients needed for matching
a low-multiplicity NLO calculation to \HWpp. They also give the
necessary and sufficient information for the matching 
of large-multiplicity processes, which we believe is best carried
out in the context of a fully automated approach to NLO cross
sections, and which we intend to pursue exploiting the work
done in ref.~\cite{Frederix:2009yq}.

\section*{Acknowledgments}
We acknowledge the collaboration of Oluseyi Latunde-Dada and Peter 
Richardson in the early stages of this work, and thank them and other 
\HWpp\ authors for helpful discussions.
B.R.W. is grateful to the CERN theory group for frequent hospitality.
The work of P.T. is supported by the Swiss National Science Foundation.

\appendix
\section{Parton showering in Herwig++\label{sec:HWppzka}}
As discussed in sect.~\ref{sec:MCsubt}, to evaluate the Monte Carlo
subtraction terms (\ref{eq:shin}) and (\ref{eq:shout}) we need
three kinds of information concerning the PSMC:
\begin{enumerate}
\item The shower variables, $z$ and $\tq$ in the case of \HWpp, and
  their expressions in terms of the variables used for the NLO
  calculation.  In general, these expressions will be different for
  initial and final state showering, as indicated by
  eqs.~(\ref{eq:shin}) and (\ref{eq:shout}) respectively.
\item The regions of phase space, if any, that are not covered by the
  parton showers at NLO (the so-called {\it dead zones}).
\item The splitting kernels used to generate the showers,
 $P_{ab}(z,\tq)$ in the case of  \HWpp.
\end{enumerate}
The showering formalism adopted in \HWpp\ is that described in
ref.~\cite{Gieseke:2003rz}.   We recall the relevant features here and
derive the necessary formulae relating the shower variables to the
invariants defined in earlier \MCN\ publications: see for example
table~4 of ref.~\cite{Frixione:2005vw}.

In \HWpp, as in all parton shower event generators, one starts with a
(usually $2\to 2$) hard subprocess and develops the external coloured legs into
jets according to the showering formalism.  The jets are then combined
according to some kinematic reconstruction method that conserves
4-momentum, preserves the internal structure of the jets and in some
sense maintains the configuration of the hard subprocess. This is not
an unambiguous procedure, and \HWpp\ includes options for different
methods.  We assume the one corresponding to the parameter values {\tt
  ReconstructionOption=General} and (for initial-state showering) {\tt
  InitialInitialBoostOption=LongTransBoost}.  These ensure that the jets, once
formed, are treated in the same way as in Fortran \HW, so the formulae 
in earlier \MCN\ papers apply once the invariant properties of a jet have 
been expressed in terms of the new shower variables.

Initial-state partons are always treated as massless in \HWpp, whereas
final states may involve partons with non-zero, possibly unequal,
masses.  We treat the more complicated case of final-state showering
first.  Then in many (but not all) respects, the initial-state case
simply corresponds to the massless limit of the same formalism.

\subsection{Final-state emission}\label{sec:fsr}
In showering from final-state partons, \HWpp\ with \MCN\ runs in a
reconstruction mode that preserves the 4-momentum of the hard
subprocess.  This is achieved by aligning the shower axes with the
directions of their parent partons in the subprocess rest frame and
then boosting the showers along their axes by appropriate amounts. In
order to be definite, we consider the case of the $2\to 2$ Born subprocess
\beq
a_+(\bp_1)+a_-(\bp_2)\;\longrightarrow\; a_{f_1}(\bk_1)+a_{f_2}(\bk_2)
\label{Rproc22}
\eeq
and the associated real-emission process
\beq
a_+(p_1)+a_-(p_2)\;\longrightarrow\; a_{f_1}(k_1)+a_{f_2}(k_2)+a_{f_3}(k_3)\;.
\label{Rproc23}
\eeq
The masses of the outgoing partons $a_{f_1}$ and $a_{f_2}$ will be
denoted by $m_1$ and $m_2$, while the initial-state partons $a_{\pm}$
and the emitted parton $a_{f_3}$ are always treated as massless.
The $2\to 2$ kinematics are defined by barred invariants:
\beqn\label{eq:bsbtbu}
\bs &=&  2\,\bp_1\cdot \bp_2 = 2\,\bk_1\cdot \bk_2+m_1^2+m_2^2\;,\nonumber\\
\bt &=& -2\,\bp_1\cdot \bk_1 =-2\,\bp_2\cdot \bk_2-m_1^2+m_2^2\;,\nonumber\\
\bu &=& -2\,\bp_1\cdot \bk_2 =-2\,\bp_2\cdot \bk_1+m_1^2-m_2^2\;,\\
0 &=& \bs+\bt+\bu\;.\nonumber
\eeqn
The definitions of the $2\to 3$ invariants used here are as in table 4 of
ref.~\cite{Frixione:2005vw}.  The relationships between the $2\to 3$ and
and $2\to 2$ invariants implied by the \HWpp\ kinematic reconstruction method
are complicated but, as mentioned above, they are the same as in \HWs,
so we also refer the reader to ref.~\cite{Frixione:2005vw} for them.

The properties of an emission from a particular external line,
say $a_{f_1}\to a_{f_1}+a_{f_3}$, that are invariant under kinematic
reconstruction are: (i) the invariant mass of the pair $(k_1+k_3)^2$,
and (ii) the `+' momentum fraction of the emitted massless parton,
\beq
\zeta_1 \equiv \frac{n\cdot k_3}{n\cdot(k_1+k_3)}\;,
\eeq
where $n$ is a light-like reference vector antiparallel to the boost
axis (i.e.\ along $a_{f_2}$ in this case).  In terms of the invariants 
defined in ref.~\cite{Frixione:2005vw}, i.e.\ $s=2\,p_1\cdot
p_2$, $w_1=2\,k_1\cdot k_3$ and $w_2=2\,k_2\cdot k_3$, we have
\beq\label{eq:jetmass}
(k_1+k_3)^2 = w_1+m_1^2
\eeq
and one can prove (see ref.~\cite{Frixione:2005vw}) that the definition 
of $\zeta_1$ corresponds to
\beq
\zeta_1=\frac{(2s-(s-w_1)\epsilon_2)w_2+(s-w_1)
[(w_1+w_2)\beta_2-\epsilon_2 w_1]}
{(s-w_1)\beta_2[2s+(s-w_1)(\beta_2-\epsilon_2)]}\;,
\eeq
with
\beqn
\epsilon_2&=&1-\frac{m_1^2-m_2^2}{s-w_1}\;,\nonumber\\
\beta_2&=&\sqrt{\epsilon_2^2-\frac{4 s m_2^2}{(s-w_1)^2}}\;.
\eeqn

For showering by a final-state parton of mass $m_1$, the shower variables 
of \HWpp\ are
\beqn\label{eq_fztk}
z &=&\frac{n\cdot k_1}{n\cdot(k_1+k_3)}=1-\zeta_1\;,\nonumber\\
\tq &=& \frac{\kt^2}{z^2(1-z)^2} +\frac{m_1^2}{z^2}\;,
\eeqn
where $\kt$ is the transverse momentum of the emission, which is
related to the invariant $w_1$ via eq.~(\ref{eq:jetmass}):
\beq
\kt^2 = (1-z)[zw_1-(1-z)m_1^2]\;,
\eeq
so that
\beq
\tq = \frac{w_1}{z(1-z)}=\frac{w_1}{\zeta_1(1-\zeta_1)}\;.
\eeq
In the case of emission from final-state parton $a_{f_2}$, one should 
simply replace $m_1\leftrightarrow m_2$ and $w_1 \leftrightarrow w_2$.

The upper limit on the variable $\tq$, which sets the initial scale
for the shower, is related to the colour connection structure of the
$2\to 2$ hard subprocess.  If the final-state partons $a_{f_1}$ and $a_{f_2}$ 
are colour
connected, as in $s$-channel single top production, the scale is set
by the c.m.\ energy squared $\bs =2\,\bp_1\cdot \bp_2$.   Since
the hard subprocess 4-momentum is preserved in final-state emission,
in this case we have $\bs =s$.
Then, as discussed in ref.~\cite{Gieseke:2003rz}, in order to yield the
correct distribution of soft gluon radiation the regions filled by emissions
from partons $a_{f_1}$ and $a_{f_2}$ should extend up to $\tq_{f_1}$ and 
$\tq_{f_2}$, respectively, where
\beq\label{eq:tkffcon}
(\tq_{f_1}-m_1^2)(\tq_{f_2}-m_2^2) = \frac 14(s-m_1^2-m_2^2+\lambda)^2
\;\;\;\mbox{(final-final colour connection),}
\eeq 
with
\beq\label{eq:lambda}
\lambda = \lambda(\sqrt s,m_1,m_2)
\equiv\sqrt{(s+m_1^2-m_2^2)^2-4sm_1^2}
      =  \sqrt{(s-m_1^2+m_2^2)^2-4sm_2^2}\;.
\eeq
The default choice, which is adopted in \MCN, is to take
\beq
\tq_{f_1} = \frac 12(s+m_1^2-m_2^2+\lambda)\;,\;\;\;
\tq_{f_2} = \frac 12(s-m_1^2+m_2^2+\lambda)\;.
\eeq
If a given value of the NLO invariants corresponds to $\tq>\tq_{\fsa}$ for
emission from parton $a_{\fsa}$ ($\alpha=1$ or 2), then that value lies
in the dead zone for emission from that parton, and the corresponding
MC subtraction term (\ref{eq:shout}) vanishes there.

The splitting kernels used in \HWpp\ include mass corrections
appropriate to the quasi-collinear limit, as derived in
ref.~\cite{Catani:2000ef}.  In terms of the above shower variables,
for emission from parton $a_{\fsa}$, this limit corresponds to $m_\alpha,
\tilde q\to 0$ with $m_\alpha/\tilde q$ finite.  For the splitting $q\to qg$, 
we then have
\beq
P_{qq}(z,\tq) = \frac{C_F}{1-z}\left[1+z^2-\frac{2m_q^2}{z\tq}\right]\;.
\eeq
Recalling from eq.~(\ref{eq_fztk}) that the collinear limit $\kt\to 0$
corresponds to $m_q/\tilde{q}\to z$, we see that emission extends down to
zero angle, so that (unlike in \HWs) there is no empty ``dead cone'' around
the collinear direction, although emission is suppressed in this region.

In the case of $g\to gg$ there are of course no mass corrections.
For completeness, we note also the quasi-collinear kernel for
$g\to q\bar q$,
\beq
P_{qg}(z,\tq) = T_R\left[1-2z(1-z)+\frac{2m_q^2}{z(1-z)\tq}\right]\;,
\eeq
although the mass correction for this splitting does not enter into
any of the \MCN\ calculations considered here.

If the colour structure of the hard subprocess is such that an
emitting final-state parton, say  $a_{f_1}$, is colour connected to an
initial-state parton, say $a_+$, the upper limit for its shower and
that of $a_+$ is set by the corresponding momentum transfer in the hard 
subprocess, given in this case by the variable $\bt$, for which the
expression in terms of $2\to 3$ invariants may be found in 
ref.~\cite{Frixione:2005vw}.
Again as discussed in ref.~\cite{Gieseke:2003rz}, in order to yield the
correct distribution of soft gluon radiation the regions filled by emissions
from partons $a_+$ and $a_{f_1}$ should extend up to $\tq_+$ and $\tq_{f_1}$,
respectively, where now
\beq\label{eq:tkifcon}
\tq_+(\tq_{f_1}-m_1^2) = (2\,\bp_1\cdot\bk_1)^2 =|\bt|^2\;\;\;
\mbox{(initial-final colour connection).}
\eeq
In \MCN\ we use the default choice $\tq_+=|\bt|, \tq_{f_1}=|\bt|+m_1^2$.

If the colour connection of $a_{f_1}$ is instead to parton $a_-$, then 
$|\bt|$ is replaced
by $2\,\bp_2\cdot\bk_1 = |\bu|+m_1^2-m_2^2$.  Similarly, if parton
$a_{f_2}$ is connected to $a_+$, we have
 \beq\label{eq:tkifcon2}
\tq_+(\tq_{f_2}-m_2^2) = (2\,\bp_1\cdot\bk_2)^2 =|\bu|^2\;,
\eeq
with the default choice $\tq_+=|\bu|, \tq_{f_2}=|\bu|+m_2^2$, while if
$a_{f_2}$ is connected to $a_-$, then $|\bu|$ is replaced
by $2\,\bp_2\cdot\bk_2 = |\bt|-m_1^2+m_2^2$. 
 
We discuss in the next subsection how the shower variable $\tq$ is
defined for initial-state emission.

\subsection{Initial-state emission}
In showering from initial-state partons, the 4-momentum of the hard
subprocess is not preserved: it recoils longitudinally and transversely
from the emissions in the showers.  \HWpp\ with \MCN\ runs in a
reconstruction mode that preserves the invariant mass and rapidity of the hard
subprocess. Considering again the $2\to 2$ hard subprocess
(\ref{Rproc22}) and the corresponding real emission (\ref{Rproc23}),
suppose that the massless parton $a_{f_3}$ is emitted from $a_+$, also
taken to be massless.  The \HWpp\ variables for an initial-state
shower are
\beq\label{eq_iztk}
z=1-\frac{n\cdot k_3}{n\cdot p_1}\;,\;\;\;
\tq = \frac{\kt^2}{(1-z)^2}\;.
\eeq
Since the two initial-state showers are (anti-)aligned and the other
incoming massless parton $a_-$ does not emit, we can take $n=p_2$.
In terms of the invariants of \cite{Frixione:2005vw}, i.e.\
$s=2\,p_1\cdot p_2$, $v_1=-2\,p_1\cdot k_3$ and $v_2=-2\,p_2\cdot k_3$, 
we then have
\beq
z = 1+\frac{v_2}{s}
\eeq
and $\kt^2 = -(1-z)v_1$, so that
\beq
\tq = \frac{v_1\,s}{v_2}\;.
\eeq
For emission from incoming parton $a_-$, interchange $v_1\leftrightarrow
v_2$.

For subprocesses where the initial-state partons $a_+$ and $a_-$ are
colour connected, the scales for their showers are set by the hard
subprocess invariant mass squared, $\bs$. 
The showers from partons $a_+$ and $a_-$ should extend up to $\tq_+$
and $\tq_-$ such that
\beq\label{eq:tkiicon}
\tq_+\tq_-=\bs^2\;\;\;\mbox{(initial-initial colour connection).}
\eeq
and we adopt the default values $\tq_+=\tq_-=\bs$.  This looks similar
to the case of final-final colour connection for massless partons, 
eq.~(\ref{eq:tkffcon}).
Note however that in the present case we have $\bs\neq s$ due to recoil 
effects; in fact
\beq
\bs = s+v_1+v_2\;.
\eeq

If on the other hand the colour connection is from $a_+$ to final-state
parton $a_{f_1}$, then we follow the prescription (\ref{eq:tkifcon}) instead
of  (\ref{eq:tkiicon}), and the upper limit on $\tq_+$ is given by
$\tq_+=|\bt|$.  If the colour connection is instead to parton
$a_{f_2}$, then the limit is $|\bu|$.  The limits for initial-final
connections of parton $a_-$ follow analogously.

Table~\ref{tab:ccon} summarizes the limits on $\tq$ for all the
possible colour connections. 
If the emitted parton $a_{f_3}$ is a light quark instead of a gluon,
as can happen in initial-state showering, for example in $gq\to Z^0q$,
the same limits apply, even though the soft gluon
radiation pattern is not relevant.  This is because \HWpp\ evolves the
shower downwards from the limiting scale and does not determine in
advance whether a quark or a gluon will be emitted in the first splitting.

%%%%%%%%%%%%%%%%%%%%%%%%%%%%%%%%%%%%%%%%%%%%%%%%%%%%%%%%%%%%%%%%%%%%%%%%%%
\begin{table}[htb]
\renewcommand{\arraystretch}{1.2}
\begin{center}
\begin{tabular}{|c|c|l|}
\hline
Parton & Col. con. & Limit \\
\hline\hline
$a_+$   &  $a_-$ & $\bs$ \\
$a_+$   &  $a_{f_1}$ & $|\bt|$ \\
$a_+$   &  $a_{f_2}$ & $|\bu|$ \\
\hline
$a_-$   &  $a_+$ & $\bs$ \\
$a_-$   &  $a_{f_1}$ & $|\bu|+m_1^2-m_2^2$ \\
$a_-$   &  $a_{f_2}$ & $|\bt|-m_1^2+m_2^2$ \\
\hline
$a_{f_1}$   &  $a_+$ & $|\bt|+m_1^2$ \\
$a_{f_1}$   &  $a_-$ & $|\bu|+2m_1^2-m_2^2$ \\
$a_{f_1}$   &  $a_{f_2}$ & $\frac 12(s+m_1^2-m_2^2+\lambda)$ \\
\hline
$a_{f_2}$   &  $a_+$ & $|\bu|+m_2^2$ \\
$a_{f_2}$   &  $a_-$ & $|\bt|-m_1^2+2m_2^2$ \\
$a_{f_2}$   &  $a_{f_1}$ & $\frac 12(s-m_1^2+m_2^2+\lambda)$ \\
\hline
\end{tabular}
\caption{\label{tab:ccon}Limits on $\tq$ for showering partons with different 
  colour connections.  The kinematic invariants are defined in
  eqs.~(\ref{eq:bsbtbu}) and (\ref{eq:lambda}).}
\end{center}
\end{table}
%%%%%%%%%%%%%%%%%%%%%%%%%%%%%%%%%%%%%%%%%%%%%%%%%%%%%%%%%%%%%%%%%%%%%%%%%% 

\section{Construction of MC subtraction terms\label{sec:MCsubt2}}
In this section we again consider the case 
of the real-emission process (\ref{Rproc23})
and label the relevant $\Sfun$ functions by
\beq
\mFKS_\pm=\Big\{(\fFKS,+),(\fFKS,-)\Big\}\,,\;\;\;\;
\mFKS_1=(\fFKS,f_1)\,,\;\;\;\;
\mFKS_2=(\fFKS,f_2)\,,
\eeq
which is more general than either eq.~(\ref{ttblabels}) or 
eq.~(\ref{stlabels}), and in fact allows one to deal with any
$2\to 3$ real-emission process (possibly after relabeling of
the partons in eq.~(\ref{Rproc23})). The subtracted real-emission
contribution to the NLO cross section 
read~\cite{Frixione:1995ms,Frederix:2009yq}
\beqn
d\hat{\sigma}_{\mu|\mFKSi}^{(3)}&=&\half\xic\left[\omyid+\opyid\right]
\Big((1-\yi^2)\xi^2\matrmu\Big)\Sfun_{\mu|\mFKSi}\, 
\nonumber \\*&\times&
d\xi d\yi d\phii d\tilde{\phi}_2^{(\mFKSi)}\,,
\label{FKSrin}
\\
d\hat{\sigma}_{\mu|\mFKSf}^{(3)}&=&\xic\omyjd
\Big((1-\yj)\xi^2\matrmu\Big)\Sfun_{\mu|\mFKSf}\, 
%\nonumber \\*&\times&
d\xi d\yj d\phij d\tilde{\phi}_2^{(\mFKSf)}\,.\phantom{aa}
\label{FKSrout}
\eeqn
The variables used in these equations are always defined in the
c.m. frame of the colliding partons. We have denoted by $\xi$ 
the energy of parton $a_{f_3}$, divided by $\sqrt{s}/2$, with 
$s$ the c.m. energy squared. In eq.~(\ref{FKSrin}), $\yi$ denotes
the cosine of the angle between partons $a_{f_3}$ and $a_+$, while
in eq.~(\ref{FKSrout}) $\yj$ denotes the cosine of the angle between
partons $a_{f_3}$ and $a_{\fsa}$. The quantities $\phii$ and $\phij$
are azimuthal angles, whose definitions are not relevant in what follows.
The plus distributions in $\xi$, $\yi$, and $\yj$ that appear in
eqs.~(\ref{FKSrin}) and~(\ref{FKSrout}) subtract the soft, initial-state
collinear, and final-state collinear singularities respectively.
By construction, the three-body phase space is
\beqn
d\phi_3&=&\xi d\xi d\yi d\phii d\tilde{\phi}_2^{(\mFKSi)}
\label{phspi}
\\*&=&
\xi d\xi d\yj d\phij d\tilde{\phi}_2^{(\mFKSf)},
\label{phspf}
\eeqn
where the measures $d\tilde{\phi}_2^{(\mFKS)}$ have the same
dimensionality of the two-body phase space $d\phi_2$, and are
proportional to it in the relevant soft and collinear limits.
Using eqs.~(\ref{phspi}) and~(\ref{phspf}), we see that the {\em un}subtracted 
real-emission cross sections (i.e., the quantities obtained from 
eqs.~(\ref{FKSrin}) and~(\ref{FKSrout}) by replacing the plus distributions 
with ordinary functions) are such that
\beqn
d\sigma_{\mu|\mFKSi}^{(3)}&=&\matrmu\,\Sfun_{\mu|\mFKSi}\,d\phi_3\,,
\\
d\sigma_{\mu|\mFKSf}^{(3)}&=&\matrmu\,\Sfun_{\mu|\mFKSf}\,d\phi_3\,.
\eeqn
Hence, one defines
\beq
d\bSigma_{\mu|\mFKSi}^{(3)}=\lum\,d\sigma_{\mu|\mFKSi}^{(3)}\,,
\;\;\;\;\;\
d\bSigma_{\mu|\mFKSf}^{(3)}=\lum\,d\sigma_{\mu|\mFKSf}^{(3)}\,,
\label{defbSigma}
\eeq
which is eq.~(\ref{bSigma}).

The equations given above can now be used for the explicit construction
of the MC subtraction terms $d\bSigma_{\mu|\mFKS}^{\rm\sss(MC)}$ used
in the \MCatNLO\ generating functional, eq.~(\ref{genFdef}). In particular,
the idea is to express the short-distance cross sections
of MC origin, eqs.~(\ref{eq:shin}) and~(\ref{eq:shout}), in the same
form as the NLO ones, eqs.~(\ref{FKSrin}) and~(\ref{FKSrout}).
Then, these cross sections are multiplied by the luminosity factors
that appear in eqs.~(\ref{eq:spl})--(\ref{eq:sfA}) to obtain the
MC subtraction terms, by analogy with eq.~(\ref{defbSigma}).

The manipulations of the MC short-distance cross section are based on
the following observations. Firstly, the variables $\xi$ and $\yi$
(or $\xi$ and $\yj$), introduced in the FKS subtraction method for
the integration of the NLO cross sections, are in one-to-one correspondence 
with the \HWpp\ shower variables $z_\pm$ and $\qtt_\pm$
(or $z_{\fsa}$ and $\qtt_{\fsa}$); hence,
the two pairs can be related by a change of variables. Secondly, the
Born-level cross sections that appear in eqs.~(\ref{eq:shin}) 
and~(\ref{eq:shout}) can be written as~\cite{Odagiri:1998ep}
\beq
\dsb_{\mu^\prime}^{(L,l)}=\mato_{\mu^\prime}^{(b;L,l)}\,d\phi_2\,,
\label{dsbprime}
\eeq
with
\beq
\mato_{\mu^\prime}^{(b;L,l)}=\frac{1}{{\cal N}_L}
\frac{D_{f\!(L,l)}}{\sum_{f^\prime}D_{f^\prime}}\mat_{\mu^\prime}^{(b)}\,,
\;\;\;\;\;\;
\lim_{N_c\to\infty}\mat_{\mu^\prime}^{(b)}=\sum_{f^\prime}D_{f^\prime}\,.
\eeq
Here, $D_{f^\prime}$ is the leading-$N_c$ contribution to the Born
matrix element squared, $\mat_{\mu^\prime}^{(b)}$, for a given 
colour flow ${f^\prime}$, with $f\!(L,l)$ the colour flow identified
by $L$ and $l$. The pre-factor ${\cal N}_L$ is equal to one or two in 
the case of a branching of a quark or a gluon line respectively (the 
latter choice is due to the fact that a gluon has two colour partners for 
a given colour flow, and one of them is chosen at random with probability 
equal to $1/2$). In eq.~(\ref{dsbprime}), $d\phi_2$ is
the two-body phase-space, whose explicit parametrization depends on
the leg that will eventually branch. It is clear, thus, that such
a phase-space can easily be related to $d\tilde{\phi}_2^{(\mFKSi)}$
or to $d\tilde{\phi}_2^{(\mFKSf)}$. 

We start by dealing with eq.~(\ref{eq:shin}), which we re-write
as follows:
\beqn
d\hat\sigma_{\mu}^{(\pm,l)}\xMCB &=& \frac{1}{2\xi}
\left[\frac{1}{1-\yi}+\frac{1}{1+\yi}\right]
\left((1-\yi^2)\xi^2
\frac{d\varsigma_{\mu}^{(\pm,l)}}{d\phi_3}\xMCBB\right)
\frac{d\phi_3}{\xi}\,,
\label{MCcntin}
\\
\frac{d\varsigma_{\mu}^{(\pm,l)}}{d\phi_3}\xMCBB&=&\frac{\as}{(2\pi)^2}\,
\frac{P_{a^\prime b^\prime}(z_\pm)}{\xi\,\qtt_\pm}
\mato_{\mu^\prime}^{(b;\pm,l)}\,\stepf_{\rm dead}^{(\pm,l)}\,
\frac{\partial(z_\pm,\qtt_\pm)}{\partial(\xi,\yi)}\,
\frac{d\phi_2}{~~d\tilde{\phi}_2^{(\mFKSi)}}\,.
\label{eq:shin2}
\eeqn
In order to be able to express the measure that appears in 
eq.~(\ref{eq:shin}) in terms of the three-body phase space $d\phi_3$,
we have inserted a trivial factor $d\phii/(2\pi)$ on its r.h.s.,
and made use of eq.~(\ref{phspi}). We can now insert the luminosity
factors. Equations~(\ref{eq:spl}) and~(\ref{eq:smn}) suggest we
define
\beqn
d\lum^{(+)}&=&\lum^{(+)}d\bx_{1i}\,d\bx_{2i}\,\equiv\,
\frac{1}{z_+}
f_a^{\Hone}(\bx_{1i}/z_+)f_b^{\Htwo}(\bx_{2i})\,
d\bx_{1i}\,d\bx_{2i}\,,
\\
d\lum^{(-)}&=&\lum^{(-)}d\bx_{1i}\,d\bx_{2i}\,\equiv\,
\frac{1}{z_-}
f_a^{\Hone}(\bx_{1i})f_b^{\Htwo}(\bx_{2i}/z_-)\,
d\bx_{1i}\,d\bx_{2i}\,.
\eeqn
As discussed in sect.~\ref{sec:MCsubt}, when matching a PSMC with an
NLO computation, variables $\bx_{1i}$ and $\bx_{2i}$ can be expressed
in terms of their analogues used in the NLO computation (which we
have denoted by $x_1$ and $x_2$ in eq.~(\ref{NLOxsec})). Hence
we can write
\beqn
d\lum^{(\pm)}&=&\lum^{(\pm)}
\frac{\partial(\bx_{1i},\bx_{2i})}{\partial(x_1,x_2)}
dx_1 dx_2\,.
\eeqn
Putting everything together, we are led to define
\beq
\frac{d\bSigma_{\mu|\mFKSi}^{\rm\sss(MC)}}{d\phi_{3}}=
\frac{\partial(\bx_{1i},\bx_{2i})}{\partial(x_1,x_2)}
\sum_{l}\left(
\lum^{(+)}\frac{d\varsigma_{\mu}^{(+,l)}}{d\phi_3}\xMCBB +
\lum^{(-)}\frac{d\varsigma_{\mu}^{(-,l)}}{d\phi_3}\xMCBB\right).
\label{MCsin}
\eeq
We explicitly point out that, as the notation suggests, the variables
$\bx_{1i}$ and $\bx_{2i}$ have the same functional form w.r.t. $x_1$
and $x_2$ regardless of whether the branching parton is $a_+$ or $a_-$,
and therefore one is able to factor out the jacobian factor in
eq.~(\ref{MCsin}). We also stress that the damping factor
\mbox{$(1-\yi^2)\xi^2$} that appears in eq.~(\ref{MCcntin}) is cancelled
by an identical factor in the denominator of that equation. The reason
for writing eq.~(\ref{MCcntin}) in that way is to make an explicit
connection with its real-emission counterpart, eq.~(\ref{FKSrin}).
In fact, in numerical codes it turns out to be convenient to define
as core functions the real-emission matrix elements, or the MC subtraction
terms, times the damping factor, for the simple reason that these quantities
are finite in the soft and collinear limits.

The treatment of final-state branchings is completely analogous to
that discussed above. We have
\beqn
d\hat\sigma_{\mu}^{(\fsa,l)}\xMCB &=& \frac{1}{\xi}\frac{1}{1-\yj}
\left((1-\yj)\xi^2
\frac{d\varsigma_{\mu}^{(\fsa,l)}}{d\phi_3}\xMCBB\right)
\frac{d\phi_3}{\xi}\,,
\label{MCcntout}
\\
\frac{d\varsigma_{\mu}^{(\fsa,l)}}{d\phi_3}\xMCBB&=&\frac{\as}{(2\pi)^2}\,
\frac{P_{a^\prime b^\prime}(z_{\fsa},\qtt_{\fsa})}
{\xi\,\qtt_{\fsa}}
\mato_{\mu^\prime}^{(b;\fsa,l)}\,\stepf_{\rm dead}^{(\fsa,l)}\,
\frac{\partial(z_{\fsa},\qtt_{\fsa})}{\partial(\xi,\yj)}\,
\frac{d\phi_2}{~~d\tilde{\phi}_2^{(\mFKSf)}}\,.
\label{eq:shout2}
\eeqn
Furthermore, from eq.~(\ref{eq:sfA}) we have
\beq
d\lum^{(\fsa)}=\lum^{(\fsa)}d\bx_{1f}\,d\bx_{2f}\,,\equiv\,
f_a^{\Hone}(\bx_{1f})f_b^{\Htwo}(\bx_{2f})\,
d\bx_{1f}\,d\bx_{2f}\,.
\eeq
As discussed in ref.~\cite{Frixione:2005vw}, in the case of final-state
branchings we have $\bx_{1f}=x_1$ and $\bx_{2f}=x_2$. Therefore
\beq
\frac{d\bSigma_{\mu|\mFKSf}^{\rm\sss(MC)}}{d\phi_{3}}=
\sum_{l}\lum^{(\fsa)}\frac{d\varsigma_{\mu}^{(\fsa,l)}}{d\phi_3}\xMCBB\,.
\label{MCsout}
\eeq

\end{document}